\documentclass[useAMS, usegraphicx,usenatbib]{mn2e}
\usepackage[utf8]{inputenc}
\usepackage[T1]{fontenc}
\usepackage{graphicx}
\usepackage[fleqn]{amsmath}
\usepackage[figuresright]{rotating}
\usepackage{booktabs}
\usepackage{mathrsfs}
\usepackage{amssymb}
\usepackage[english]{babel}
\usepackage{amsfonts}
\usepackage{txfonts}
\usepackage{textcomp}
\usepackage{microtype}
\usepackage{aas_macros}
\usepackage[pdftex]{lscape}
\usepackage[usenames,dvipsnames]{xcolor}
\usepackage{subfig}
\usepackage{xspace}
\usepackage{my_def}
\usepackage{enumitem}
\usepackage{float}
\usepackage{aecompl}

\newcommand*{\rc}{r_\mathrm{c}}
\newcommand*{\racc}{r_\mathrm{acc}}
\newcommand*{\rmin}{r_\mathrm{min}}
\newcommand*{\Min}{\dot{M}_\mathrm{in}}
\newcommand*{\Mout}{\dot{M}_\mathrm{out}}
\newcommand*{\Lw}{L_\mathrm{w}}
\newcommand*{\vw}{v_\mathrm{w}}
\newcommand*{\pdot}{\dot{p}_\mathrm{w}}
\newcommand*{\epsem}{\epsilon_\mathrm{EM}}
\renewcommand*{\[}{\begin{equation}}
\renewcommand*{\]}{\end{equation}}
\newcommand*{\thetamax}{\theta_1}
\newcommand*{\sigtheta}{\sigma_{\theta}}
\newcommand*{\sigr}{\sigma_r}
\newcommand*{\rmiin}{r_1}
\newcommand*{\rmax}{r_2}
\newcommand*{\rb}{r_\mathrm{B}}
\newcommand*{\cs}{c_\mathrm{s}}
\newcommand*{\Mdotbondi}{\dot{M}_\mathrm{B}}
\newcommand*{\cl}{\_C\_L\xspace}
\newcommand*{\ch}{\_C\_H\xspace}

\newcommand*{\mSFR}{\left\langle\mathrm{SFR}\right\rangle}

\newcommand*{\AN}[1]{#1}

\bibpunct{(}{)}{;}{a}{}{,}

\date{Draft, \today}

\title[Black hole feeding and feedback]{Black hole feeding and feedback: the physics inside the ``subgrid''}
 \author[A. Negri, M. Volonteri]{A. Negri\thanks{E-mail:
negri@iap.fr}, M. Volonteri
\\Sorbonne Universités, UPMC Univ Paris 6 et CNRS, UMR 7095, Institut d’Astrophysique de Paris, 98 bis bd Arago, 75014 Paris, France}

\begin{document}
\maketitle
\label{firstpage}

\begin{abstract}
Black holes (BHs) are believed to be a key ingredient of galaxy formation. However, the galaxy-BH interplay is challenging to study due to the large dynamical range and complex physics involved. As a consequence, hydrodynamical cosmological simulations normally adopt sub-grid models to track the unresolved physical processes, in particular BH accretion; usually the spatial scale where the BH dominates the hydrodynamical processes (the Bondi radius) is unresolved, and an approximate Bondi-Hoyle accretion rate is used to estimate the growth of the BH. By comparing hydrodynamical simulations at different resolutions (300, 30, 3 pc) using a Bondi-Hoyle approximation to sub-parsec runs with non-parametrized accretion, our aim is to probe how well an approximated Bondi accretion is able to capture the BH accretion physics and the subsequent feedback on the galaxy. We analyse an isolated galaxy simulation that includes cooling, star formation, Type Ia and Type II supernovae, BH accretion and active galactic nuclei feedback (radiation pressure, Compton heating/cooling) where mass, momentum, and energy are deposited in the interstellar medium through conical winds. We find that on average the approximated Bondi formalism can lead to both over- and under-estimations of the BH growth, depending on resolution and on how the variables entering into the Bondi-Hoyle formalism are calculated.
\end{abstract}


\section{Introduction}
Black holes (BHs) are believed to play a crucial role in galaxy formation and evolution. The correlations between BH mass and host galaxy properties \citep[e.g.,][]{magorrian1998, ferrarese.merrit2000, gebhardt.etal2000,gultekin.etal2009, kormendy.ho2013,mcconnell.ma2013, kormendy2016}  suggest that their respective evolutions are closely tied. However, two decades after the pioneering work of \citet{magorrian1998} the impact of BHs formation and growth on the host galaxies evolution is not fully understood yet.

The BH-galaxy coevolution is believed to rely on active galactic nuclei (AGN) feedback, the prodigious energetic emission from the BH accretion disc that, from millipc scales, is able to influence the surrounding medium up to kpc scales \citep[for a review see][]{fabian2012}. The mutual interaction between BHs and their host galaxies, and in particular the BH accretion processes, has been studied both analytically and numerically. Various models have been proposed to describe BH accretion physics and the relative feedback \citep{silk.rees1998, fabian1999, king2005}. Although they have been successful in explaining, for instance, the relationship between BH mass and the galaxy velocity dispersion, $\Mbh$--$\sigma$, analytical models are forced to assume simplified geometries and neglect important physical processes such as mass inflow and star formation.

In the last 10 years hydrodynamical cosmological simulations including BH growth and feedback have become an invaluable tool to shed light on galaxy formation and evolution \citep[e.g.,][]{dimatteo.etal2005, springel.etal2005,robertson.etal2006, sijacki.etal2007, dimatteo.etal2008, booth.schaye2009,  schaye.etal2010, debhur.etal2011, degraf.etal2011, teyssier.etal2011, degraf.etal2012, dubois.etal2014, hirschmann.etal2014, vogelsberger.etal2014b, khandai.etal2015, feng.etal2016, volonteri.etal2016}. These simulations are often at the edge of modern supercomputers computational capabilities, and they have to overcome numerous computational challenges. Perhaps the most important is the involved spatial range: a proper simulation of both the large scale structure of the Universe and the BH accretion scales would require a dynamical  range spanning from the Schwarzschild radius of a super massive BH -- $r_\mathrm{s}\approx 10^{-6}~\pc$ for a mass of $10^7~\Msun$ -- to a scale of hundreds of Mpc. Bridging all the involved scales is currently computationally infeasible, and it will remain so for the near future. As a consequence, cosmological simulations including AGN feedback have to resort to so-called sub-grid models in the case of BH accretion and feedback.

Perhaps the most commonly used sub-grid accretion models are based on the Bondi-Hoyle-Lyttleton accretion solution \citep{hoyle.lyttleton1939, bondi.hoyle1944, Bondi1952},  first proposed by \citet{springel.etal2005} and then widely adopted in the literature. The main reason of its popularity relies in its simplicity: the Bondi problem describes the spherical stationary inflow of a perfect, non viscous, non self-gravitating gas on a BH, subject to polytropic transformations and in absence of any type of feedback. Given these hypotheses, the BH accretion rate can be written as
\[
\Mdotbondi = \dfrac{4\upi \lambda_\mathrm{c} G^2  \Mbh^2 \rho_\infty }{{\cs}_\infty^3}, \label{eq:bondi}
\]
where
\[\lambda_\mathrm{c} = \dfrac{1}{4} \left( \dfrac{2}{5-3\gamma} \right)^{\frac{5-3\gamma}{2\left(\gamma-1 \right)}}, \]
$G$ is the gravitational constant, $\Mbh$ is the BH mass, $\rho_\infty$ and ${\cs}_\infty$ are the density and sound speed at great distance (formally infinite) from the BH, and 
$\gamma$ is the polytropic transformation index \citep{Bondi1952}. Thus, with a simple formula  BH mass accretion can be quantified.

However, Bondi accretion suffers from various problems if applied to realistic environments. First, the accretion of a turbulent, non-uniform gas flow is not considered \citep[see][and reference therein]{barai+11, barai+12, park.ricotti2012,gaspari.etal2013}. Secondly, Bondi accretion spherical symmetry implies that gas angular momentum is negligible, which is unlikely to be true in galaxy formation simulations. Gas with angular momentum will circularize before reaching the innermost stable orbit around the BH, creating an accretion disc. Extensive numerical works addressed this issue, by either analytically modelling the angular momentum impact on accretion \citep{hopkins.quataert2011, rosas-Guevara.etal2015,tremmel.etal2016} or by directly resolving the rotating accretion flow on small scales \citep{debuhr.etal2010, power.etal2011, park.ricotti2012, li.etal2013, gaspari.etal2015, angles-Alcazar.etal2016, curtis.sijacki2016, hopkins.etal2016}. Moreover, BHs are not isolated objects as assumed in the Bondi problem, but they usually reside at the bottom of the galactic potential well. The additional pull due to the galactic gravitational field could provide additional material to the BH with respect to the standard Bondi solution, where this additional contribution is neglected \citep[for a fully analytical extended Bondi solution considering the halo gravitational pull see][]{korol.etal2016}.

Another level of complexity is then added by further physical mechanisms not considered in the Bondi solution, above all gas cooling, AGN feedback and star formation. Even evaluating the rather simple  right-hand-side of equation~(\ref{eq:bondi}) in a numerical simulation is not trivial: where should ${\rho}_\infty$ and ${\cs}_\infty$ be calculated and in which way? In addition, a comparison between Eulerian (grid) and Lagrangian (smooth particle hydrodynamics, hereafter SPH) codes is far from being trivial due to the structural difference in the fluid discretization. Finally, from the numerical point of view, Bondi accretion suffers from resolution dependent issues (see Section~\ref{codes}).

Despite the intrinsic difficulty of the Bondi approach in estimating realistic accretion rates in a non-ideal environment, it has been adopted by a large part of the numerical simulation community, providing results that broadly agree with observational constraints on BH growth and AGN evolution. Thus, in the next Section we summarize the main sub-grid processes based on Bondi accretion and their pitfalls. 

\subsection{Black hole accretion in hydrodynamical cosmological simulations}\label{codes}
All  hydrodynamical cosmological simulations employ a sub-grid recipe for BH accretion. The first implementation of BH accretion based on the Bondi solution in a hydrodynamical galaxy-scale simulation has been proposed by \citet{springel.etal2005}. The BH accretion rate was computed as $\Mdotbh = \alpha \Mdotbondi$, where $\alpha$ is a dimensionless constant named boost parameter with a typical value of 100\footnote{The value of $\alpha$ is not explicitly mentioned in the earlier works of hydrodynamical cosmological simulations, but see Table~2 of \citet{booth.schaye2009} for a list of the adopted boost parameters.}, and $\rho_\infty$, ${\cs}_\infty$ and $v_\mathrm{BH}$ were obtained using a kernel-weighted SPH estimator. The accretion rate is then capped at the Eddington limit, thus super-Eddington accretion is not allowed by construction\footnote{We do not further discuss here the validity of capping the mass accretion rate, see \citet{volonteri.etal2015} and \citet{inayoshi.etal2015} for an overview. Note that \citet{lusso.ciotti2011} describe the transition from the pure hydrodynamical to the optically thin  radiation-dominated regime by self-consistently solving the Bondi problem including electron scattering.}. This is the accretion rate employed by the Magneticum simulation \citep{hirschmann.etal2014}.

\citet{booth.schaye2009} motivate the adoption of the boosted accretion by noting that at the resolutions of cosmological simulations, the Bondi radius (the scale where the BH dominates the hydrodynamics) and the properties of the interstellar medium (ISM) cold phase are not resolved. Thus they argue that the temperature of the material accreted on the BH would be overestimated, and its density underestimated, leading to a very low accretion rate in equation~(\ref{eq:bondi}) if large values of $\alpha$ are not adopted. However, they also discuss the limits of these justifications: when the BH is in a state of a hot accretion, which typically shows small deviations from spherical symmetry, the Bondi assumptions are fulfilled, and a large boost factor is no longer needed. \AN{Moreover, the combination of the Eddington limiter and the costant $\alpha$ factor can present unexpected consequences. They show that by assuming $\alpha =100$, BHs with a mass $>10^6~\Msun$ reach a sub-Eddington accretion only when the gas density has become extremely low, almost two order of magnitude less than the star formation threshold. In other words,  BH accretion a constant boosting will be sensitive to the ISM density (and thus more easily regulated by AGN feedback) only when the BH surroundings have been almost completely emptied of gas.} To minimize these effects, they proposed a density-dependant boost factor, where $\alpha=1$ if $n_\mathrm{gas}/n_\mathrm{sf}<1$ and $\alpha=(n_\mathrm{gas}/n_\mathrm{sf})^\beta$ otherwise, where $n_\mathrm{gas}$ is the gas number density, $n_\mathrm{sf}=0.1~\cc$ is the gas density threshold for star formation and $\beta=2$; the accretion is then limited at the Eddington luminosity, assuming a radiative efficiency of 10 per cent.

Following \citet{springel.etal2005} and \citet{booth.schaye2009}, the boosted Bondi accretion has been adopted (and modified) by a large number of authors. \citet{pelupessy.etal2007} do not follow a constant $\alpha$ prescription; instead $\Mdotbondi$ is the sum of both the hot and cold phase Bondi accretion rates, calculated with boosting factors equal to their respective mass fractions. This scheme is adopted by the Massiveblack \citep{degraf.etal2012}, Massiveblack II \citep{khandai.etal2015} and BLUETIDES simulations \citep{feng.etal2016, dimatteo.etal2016}.
A similar approach is employed by \citet{steinborn.etal2015}, where they calculate the physical quantities of equation~(\ref{eq:bondi}) using 295 neighbours, and calculating separately the accretion rate for the cold and the hot gas, with $\alpha=100$ and 10, respectively (this is the default BH sink algorithm of the \textsc{gadget3-x} code). Finally, \citet{choi.etal2012} apply the average directly on $\Mdotbh$ with $\alpha=100$ calculated over 64 neighbours, in order to reduce the dependence on the numerical kernel. \AN{The actual accretion of the particles around the BH is performed in a stochastic fashion. Only the particles inside the Bondi radius are accreted, with a probability proportional to the amount of overlap between the particle and the Bondi radius (if the Bondi radius is unresolved, it is set equal to the particles softening length). An additional probability taking into account the free fall time for each particles is considered. The same prescription is used in \citet{choi.etal2015} without $\alpha$ boosting.}

Some of these prescriptions have been adopted as the default accretion algorithm in numerical codes. The \textsc{gadget3-owls} code implements the \citet{booth.schaye2009} accretion. This modified version of GADGET 3 is employed in the \textsc{owls} \citep{schaye.etal2010}, \textsc{cosmo-owls} \citep{lebrun.etal2014} and BAHAMAS simulations \citep{mcCarthy.etal2016}. The prescription has been modified in the EAGLE simulation \citep{schaye.etal2015} to take into account \AN{the effects of gas angular momentum on BH accretion \citep[adopted from][]{rosas-Guevara.etal2015}; at variance with the majority of cosmological simulations, the boost factor is no longer employed.}

The implementation in \textsc{changa} modifies the Bondi formalism used in \textsc{gasoline} \citep{bellovary.etal2010} to include the suppressive effect of gas rotational support \citep{tremmel.etal2016}. In both cases the accretion rate is computed as the sum of the Bondi accretion rate of each of the 32 individual gas particles nearest the BH, rather than simply averaging the gas quantities to insert them in equation~\ref{eq:bondi}.  A boost factor may or may not be included. 

BHs in the \textsc{arepo} code \citep{springel2010} are collision-less sink particles whose accretion rate is calculated following equation~(\ref{eq:bondi}), where a volume weighted mean is performed \citep{vogelsberger.etal2013, sijacki.etal2015}. However, \citet{vogelsberger.etal2013} note that when no star-forming gas is present in the immediate vicinity of the BH, the boosted Bondi prescription can lead to an overestimate of the mass accretion rate. Thus, in this regime they lower the accretion rate based on a pressure criterion. This configuration has been employed in the Illustris simulation \citep{vogelsberger.etal2014b}, while \citet{curtis.etal2015} proposed a new super-Lagrangian refinement to reach the Bondi radius near a BH.

Moving to adaptive mesh refinement, in \textsc{ramses} \citep{teyssier2002} BHs accrete following equation~(\ref{eq:bondi}) where the right-hand-side is composed by mean values calculated by disseminating clouds particles around the BH then weighted using a kernel \citep{dubois.etal2012}. The accretion rate is then boosted following the \citet{booth.schaye2009} prescription. In the Horizon-AGN simulation \citep{dubois.etal2014}, the mean quantities are computed in a mass weighted fashion (Dubois, private communication). New sink schemes are proposed in \citet{bleuler.teyssier2014}. All the above mentioned sub-grid models are Eddington limited. In the  \textsc{enzo} code \citep{kim.etal2011, bryan.etal2014} Bondi accretion is assumed, and the density in equation~(\ref{eq:bondi}) is extrapolated at the Bondi radius, no boost factor is taken into account.

Despite the details of the BH accretion scheme not always being stated explicitly in the literature, two clear approaches emerge: volume weighting (the SPH kernel weighted estimators are similar to volume weighting) and mass weighting for the calculation of the Bondi rate. Finally, there have been recent proposals that forgo the Bondi approximation in favour of angular-momentum driven sub-grid implementations \citep[e.g. see][]{debuhr.etal2010, angles-Alcazar.etal2016}.

All the sub-grid algorithms presented above rely on the common assumption that at the resolutions of cosmological simulations the ISM in the surroundings of a BH is smoothed, so its density and temperature are respectively under and overestimated, leading to an accretion rate lower than in reality. However, the boost factor $\alpha$ is poorly constrained, being problem and resolution dependent. In addition, comparative studies such as \citet{wurster.thacker2013} and \citet{elahi.etal2016} employ BH sub-grid recipes in their simulations, while a comparison with a non-parametrized accretion would be ideal. High resolution studies, where the Bondi radius is resolved and the accretion rate is computed in a non-parametrized way, only simulate the innermost regions of a galaxy \citep[up to the first few hundred of pc, see][]{kurosawa.proga2009a, kurosawa.proga2009b, kurosawa.etal2009, barai+11, barai+12, hopkins.quataert2011, hopkins.etal2016}, or multiple simulations on different scales of a single object \citep{hopkins.quataert2010}. Multi-scale studies in presence of feedback has been performed by Ciotti, Ostriker, and collaborators \citep[][and references therein]{ciotti.ostriker1997, ciotti.ostriker2001, ciotti.ostriker2007, ostriker.etal2010, novak.etal.2011, ciotti.etal2016}, but no comparison with Bondi accretion has been performed. One should not be surprised in finding that at a particular time the $\Mdotbondi$ under- or over-estimate the real accretion rate, e.g. when large amounts of gas very far from the BH produces a large $\Mdotbondi$, while the BH can actually be in a state of low accretion. However, due to the non-linear nature of hydrodynamics and in presence of additional physical processes such as radiative cooling, AGN feedback and star formation it is extremely difficult to quantify a priori both the BH growth with time and the global effect of adopting the Bondi solution as a proxy for BH accretion.

Our aim is to probe two types of Bondi algorithms (using mass and volume weighted quantities) and their global effects on the BH mass by comparing 2.5D hydrodynamical simulations of an isolated galaxy at different resolutions to sub-parsec runs with non-parametrized accretion (calculated as the mass flux on the BH, see Section~\ref{bh_accretion}). In our reference run we simulate radii from 0.1~pc to 250~kpc, thus resolving the Bondi radius for the adopted BH mass at all the possible gas temperatures (Section~\ref{sec:grid}), in presence of cooling, star formation, Type Ia and Type II supernovae, and AGN feedback (radiation pressure, Compton heating/cooling, broad absorption line winds). 

The paper is organized as follows: in Section~\ref{sec:simulations} we describe the code input physics and our initial conditions. In Section~\ref{results} we present the results of our simulations, while in Section~\ref{sec:discussion} we discuss our results and we compare them to other recent works. Finally, in Section~\ref{sec:conclusions} we draw our conclusions.

\section{The simulations}\label{sec:simulations}
We use a customized, parallel version of the  hydrodynamical code \textsc{zeus} \citep{stone.norman.1992, hayesetal2006}. For the most part the detailed input physics is described in \citet[hereafter N11 and N12]{novak.etal.2011, novak.etal2012}, here we briefly summarize the main implementations and differences with respect to N11 and N12.

\subsection{The input physics}\label{input_physics}
The mass, momentum and energy injections/sinks from stellar evolution, Type Ia and II supernovae, star formation and radiative cooling (bremsstrahlung and lines cooling, with an exponential cut-off below $10^4$~K) are the same as N11. In particular, the star formation recipe can reproduce the Kennicutt--Schmidt relation \citep{kennicutt1998} with an adopted efficiency equal to 0.1.

Radiative feedback from the BH is taken into account by considering photo-ionization heating, Compton heating/cooling in the energy equation, and radiation pressure due to electron scattering and photo-ionization in the momentum equation. Given a BH accretion rate $\Mdotbh$, the total BH luminosity is:
\[
 \Lbh = \epsem \Mdotbh c^2,
\]
where $c$ is the speed of light, and $\epsem$ is the electromagnetic efficiency. We here adopt a dependence of  $\epsem$ on the accretion rate that captures the transition from optically thick and geometrically thin, radiatively efficient accretion discs \citep{Shakura.Sunyaev73}, to optically thin, geometrically thick, radiatively inefficient advection dominated accretion flows \citep{narayan.yi1994}:
\[
 \epsem =  \dfrac{\epsilon_0 A \mdot}{1+A\mdot}, \qquad \mdot\equiv \dfrac{\Mdotbh}{\Mdotedd} = \dfrac{\epsilon_0 \Mdotbh c^2}{\Ledd},
 \label{eq:epsem}
\]
here $\mdot$, $\Mdotedd$ and $\Ledd$ are the dimensionless BH and Eddington mass accretion rates, respectively, with $A=100$ and $\epsilon_0=0.125$, while $\Ledd = 4\upi G \Mbh \mp c/\sigma_\mathrm{T}$ is the Eddington luminosity, where $\mp$ and $\sigma_\mathrm{T}$ are the proton mass and the Thomson cross section.

Mechanical feedback is included as a broad absorption line (BAL) wind originating from the BH accretion disc. In N11, the BAL wind is modelled as a \AN{kinetic wind} following \citet{ostriker.etal2010}, which guarantees that the mass, energy and momentum carried by the wind are self-consistent. In this approach, the mass outflow \AN{(towards the BH)} at the inner radial grid is transferred into a sub-grid accretion disc model, then a fraction of the inflow is injected back into the computational grid as a BAL wind, while the rest is accreted on the BH. This model can be formalized as:
\begin{gather} 
 \Mdotbh = \dfrac{\Min}{1+\eta},\\[2ex]
 \Mout=\eta \Mdotbh,\label{eq:Mout}\\[2ex]
\Lw = \epsw\Mdotbh c^2,\label{eq:Lwind}\\[2ex]
 \pdot = \eta\Mdotbh \vw,\label{eq:Pwind}\\[2ex]
 \eta\equiv 2 \epsw c^2 / \vw^2,\\[2ex]
\epsw =  \dfrac{{\epsw}_0 A_w \mdot}{1+A_w\mdot}, \label{eq:epsw}
\end{gather}
where $\Min$ is the mass inflow rate at the innermost radial boundary, $\Mout$, $\epsw$, $\Lw$, $\pdot$ and $\vw$ are the BAL wind injection mass rate, and efficiency, mechanical luminosity, injection momentum rate, injection velocity, respectively; in this work we consider $\vw=10^{4}~\kms$, ${\epsw}_0=10^{-4}$ and $A_w=1000$.

\AN{We model the BAL wind as a direct injection, at the first radial grid, of mass and momentum within two symmetric cones above and below the equatorial plane, with a fixed half-opening angle of $45\degr$ containing half of the energy emitted, i.e., this means that the wind is visible from $\sim 1/5$ of the entire solid angle. Equations~(\ref{eq:Mout})-(\ref{eq:Pwind}) provide the total injection rates, which are distributed on the innermost radial grid following the prescription of \citet{ciotti.etal2016} (see their equations (25)-(28)). By defining the wind thermalised 
when its velocity drops below 5 times the velocity dispersion of the halo (150 km/s), in the region outside the double cone with semi-aperture of 45 degrees, the thermalisation radius is typically $\sim$10 pc, with a maximum at 40 pc, while in the polar region is $\lesssim$100 pc, with maximum of 300 pc.}

\AN{We stress that AGN feedback is a combination of radiative  (via photo-ionization heating and Compton heating/cooling) and kinetic forms, instead of the thermal feedback (injection of thermal energy energy with an  efficiency parameter setting the radiation-gas coupling) usually employed in cosmological simulations \citep[see also][]{barai.etal2014}.}
At variance with N12 in this work we do not consider the treatment of dust formation and destruction, and the associated radiation pressure effects \citep[see also][]{hensley.etal2014}, neither radiation pressure due to stellar light in different bands \citep[e.g., see][]{ciotti.Ostriker.2012}. 
\subsection{Numerical grids}\label{sec:grid}
The code employs a 2D axisymmetric, computational grid in spherical coordinates $(r,~\theta)$. The mesh is not dynamically refined, however the non-uniform spacing allows to concentrate more gridpoints near the centre of the simulations box. \AN{2D axisymmetric simulations offer the best tradeoff between realism and computational requirements. Neglecting the dependence from the azimuthal angle drastically decreases the computational effort; at the same time, Rayleigh-Taylor instabilities can fragment a cold shell, thus forming a clumpy, turbulent multiphase medium, and the BH can simultaneously accrete and eject mass in a non-spherical shape, which is not allowed in 1D runs. However, in absence of non-axisymmetric torques an infalling rotating gas blob will circularize without feeding the BH. For this very reason, we neglect a large level of rotation, in order to accrete on the central object.}

In this work we adopt different radial spacing for different sets of simulations, whereas the $\theta$ setup is maintained fixed. For our reference models (see Sections~\ref{bh_accretion} and \ref{sec:flux}), the mesh consists of $512 \times 30$ gridpoints, spanning from 0.1~pc to 250~kpc in the radial direction (with the spacing growing following a geometric progression), with a central resolution of 0.1~pc. The grid in the $\theta$ direction is regularly sampled and to avoid the coordinate singularity on the axis of symmetry we exclude a narrow double-cone of semi-aperture $\upi \e{-4}$~rad. In this configuration, we obtain an almost constant 0.1~pc resolution in the first 10~pc from the centre, and then resolution grows linearly with radius. As mentioned above, some simulation sets adopt a different, low resolution grid (3, 30, 300~pc), while the innermost and outermost radial grid are maintained at 0.1~pc and 250~kpc, respectively.

In hydrodynamical cosmological simulations BH accretion is commonly treated as sub-grid physics. Due to the large dynamical range required to resolve properly galaxy evolution in a cosmological context, one has to account for extremely different spatial scales. A crucial scale for the galaxy-BH co-evolution is the Bondi radius, formally defined as $\rb \equiv G\Mbh / \cs^2$, where the BH dominates the hydrodynamical evolution of the ISM. The temperature dependence is apparent: given $\Mbh$, the higher the gas temperature, the smaller the Bondi radius. Thus a numerical simulation that aims at following the ISM during accretion events should properly resolve the Bondi radius for all the temperatures of the inflowing gas. In our case, with the high-resolution grid we resolve spatial scales corresponding to temperatures less than $10^{8}$~K. Note that while the highest temperature allowed in the code derives from the  complete thermalization of the BAL wind ($\simeq 2.5\e{9}$~K), this hot gas will be outflowing by construction. As a consequence, the temperature upper limit for the inflowing gas is the Compton temperature $T_\mathrm{c} = 2\e{7}$~K, so we actually resolve the Bondi radius for all the (inflowing gas) temperatures.

For all the simulations in this work, reflecting boundary conditions were set along the $\theta$ boundary (with the gas rotational velocity flipped in sign); at the outer edge of the spherical simulated box the fluid is free to flow out of the computational grid. The inner boundary conditions are a special case: they are set on reflection when we adopt Bondi accretion (see Section~\ref{bh_accretion}), while in the case of non-parametrized accretion we adopt outflow conditions (off of the computational grid towards the centre of the simulation) reported in N11. As already noted in \citet{prasad.etal.2015}, the cold gas can stick to the $\theta$ boundary when reflecting boundary conditions are applied, thus inducing an unphysical polar accretion flow. To avoid this phenomenon, they exclude the first 8 grid zones near each pole in the calculation of the BH accretion rate. However, this can still lead to an enhanced star formation confined in the polar direction due to the cold gas presence, therefore we tuned the initial ISM angular momentum distributions near the poles to minimize cold ISM accumulation along the $\theta$ boundary (see below).
%
%
%
%
%
%
\subsection{BH accretion}\label{bh_accretion}
We employ different types of accretion models. For our high resolution, reference simulations, we employ a non-parametrized accretion scheme, ``flux" hereafter: the mass accretion rate $\Min$ is simply the flux out of the grid and towards the BH calculated on the innermost radial grid, which is at a distance of 0.1~pc from the BH. This allows the code to treat the ISM hydrodynamics self-consistently, since the gas is accreted on the BH if and only if is actually crossing our innermost cell, well below the Bondi radius.

To quantitatively study the effect of the most common sub-grid accretion prescriptions adopted in hydrodynamical cosmological simulations, we implemented four different types of Bondi accretion schemes. For each scheme we define an accretion radius $\racc$, that includes the volume where we calculate $\rho_\infty$ and ${\cs}_\infty$ in equation~(\ref{eq:bondi}) as a volume or mass weighted average. Two accretion schemes employ $\Min = \Mdotbondi$, and other two cap accretion at the accretion rate corresponding to the Eddington luminosity assuming a radiative efficiency of 10\%, i.e., $\Min = \min (\Mdotbondi, \Mdotedd)$. Thus, the four schemes differ in the way $\rho_\infty$ and ${\cs}_\infty$ are calculated and whether an Eddington limit is applied. 
In order to ensure mass conservation, when the BH accretes a certain mass of gas $M$ in a timestep, we drain the same amount of gas from the volume inside $\racc$. For each cell the rate of mass depletion $\dot{M}$ is given by: 
\(
 \dot{M}\equiv \Min M_\mathrm{cell}/M_\mathrm{tot},
\)
where $M_\mathrm{cell}$ and $M_\mathrm{tot}$ are the gas mass inside a cell and the gas mass enclosed by $\racc$, respectively. For sake of brevity from now on we will assume that a Bondi simulation employs mass weighting, when not explicitly stated.
%
%
%
%
\begin{table*}
    \caption{Mass weighted simulations.}\label{tab1}
\makebox[\linewidth]{%
\begin{tabular}{ccccccccc}
\toprule
Accretion               & Resolution & $\racc$    &   $n_0$    &  $\Delta\Mbh$ & $\Delta M_*$&   $\mSFR$ &     $\Mgas$ \\
--                      &     (pc)   &       (pc) &  ($\cc$)   & ($10^6~\Msun$)&($10^6~\Msun$)&($\Msunyr$)& ($10^{10}~\Msun$)\\
  (1)                   &      (2)   &        (3) &    (4)     &   (5)         &     (6)     & (7)       &  (8)         \\
\midrule
B300\_S\_H$^\dag$       &     0.1    &    300     &      1     &       2772.13 &      418.17 &     10.06 &      36.10  \\
B30\_S\_H               &     0.1    &     30     &      1     &         15.52 &     2092.68 &     41.85 &      35.99  \\
B3\_S\_H                &     0.1    &      3     &      1     &          2.27 &     2841.47 &     56.83 &      36.84  \\
B300\_S\_L              &     0.1    &    300     &      0.01  &         12.86 &        1.05 &      0.02 &       5.39  \\
B30\_S\_L               &     0.1    &     30     &      0.01  &          8.86 &        4.68 &      0.10 &       5.39  \\
B3\_S\_L                &     0.1    &      3     &      0.01  &          1.30 &       16.71 &      0.33 &       0.37  \\
\midrule                                                                                                             
B300\_C\_H              &    300     &    300     &      1     &        171.89 &      186.10 &      3.72 &      30.96 \\
B30\_C\_H               &     30     &     30     &      1     &        100.22 &     1085.32 &     21.71 &      31.18 \\
B3\_C\_H                &      3     &      3     &      1     &          1.24 &     3059.29 &     61.19 &      31.98 \\
B300\_C\_L              &    300     &    300     &      0.01  &         17.87 &        0.44 &      0.01 &       0.38  \\
B30\_C\_L               &     30     &     30     &      0.01  &          2.62 &        0.30 &      0.01 &       0.38  \\
B3\_C\_L                &      3     &      3     &      0.01  &          0.56 &       12.22 &      0.24 &       0.37  \\
\midrule                                                                                                             
\midrule                                                                                                             
BE300\_S\_H             &      0.1   &    300     &      1     &         32.01 &     1157.94 &     23.16 &      35.98  \\
BE30\_S\_H              &      0.1   &     30     &      1     &          7.47 &     2240.76 &     44.82 &      36.76  \\
BE3\_S\_H               &      0.1   &      3     &      1     &          1.44 &     2121.02 &     42.42 &      36.63  \\
BE300\_S\_L             &      0.1   &    300     &      0.01  &         31.63 &        1.17 &      0.02 &       5.39  \\
BE30\_S\_L              &      0.1   &     30     &      0.01  &          5.39 &        4.96 &      0.10 &       0.37  \\
BE3\_S\_L               &      0.1   &      3     &      0.01  &          1.15 &       15.41 &      0.31 &       5.38  \\
\midrule                                                                                                             
BE300\_C\_H             &    300     &    300     &      1     &         30.67 &     1787.89 &     35.76 &      30.42  \\
BE30\_C\_H              &     30     &     30     &      1     &         32.25 &     2180.58 &     43.61 &      30.70  \\
BE3\_C\_H               &      3     &      3     &      1     &          0.76 &     1103.63 &     22.07 &      30.00  \\
BE300\_C\_L             &    300     &    300     &      0.01  &         17.05 &        1.09 &      0.02 &       0.38  \\
BE30\_C\_L              &     30     &     30     &      0.01  &          1.87 &        0.35 &      0.01 &       0.38  \\
BE3\_C\_L               &      3     &      3     &      0.01  &          0.56 &       24.04 &      0.48 &       0.37  \\
\bottomrule
\end{tabular}%
}%
\flushleft
\parbox{0.9\linewidth}{\footnotesize 
    \textit{Notes.} (1) Simulation name, containing the accretion scheme (F, B, BE and BV for flux, Bondi, Bondi capped at Eddington rate and Bondi volume weighted, respectively), accretion radius (3, 30 and 300~pc), grid type (S, C for standard high resolution and coarse grid, respectively), and central initial gas number density (H or L for $n_0=1,$ $0.01~\cc$, respectively). (2) Grid resolution. (3) Accretion radius for B and BE simulations.  (4) Initial gas central number density. (5) Gas mass accreted on the BH. (6) Stellar mass formed during the run. (7) Mean SFR \AN{calculated over the entire run}. (8) Final gas mass.\\
$^\dag$ Run terminated at $t=36~\Myr$.}
\end{table*}

\begin{table*}
    \caption{Flux simulations.}\label{tabflux}
\makebox[\linewidth]{%
\begin{tabular}{ccccccccc}
\toprule
Accretion               & Resolution & $\racc$    &   $n_0$    &  $\Delta\Mbh$ & $\Delta M_*$&   $\mSFR$ &     $\Mgas$ \\
--                      &     (pc)   &       (pc) &  ($\cc$)   & ($10^6~\Msun$)&($10^6~\Msun$)&($\Msunyr$)& ($10^{10}~\Msun$)\\
  (1)                   &      (2)   &        (3) &    (4)     &   (5)         &     (6)     & (7)       &  (8)         \\
\midrule
F\_S\_H                 &      0.1   &    --      &      1     &          0.53 &     2523.00 &     50.46 &      36.66 \\ 
F\_S\_L                 &      0.1   &    --      &      0.01  &          0.49 &       14.60 &      0.29 &       0.36 \\
\bottomrule
\end{tabular}%
}%
\flushleft
\parbox{0.9\linewidth}{\footnotesize 
\textit{Notes.} Same as Table~\ref{tab1}.}
\end{table*}

\begin{table*}
\caption{Simulations without AGN feedback.}\label{tab2}
\makebox[\linewidth]{%
\begin{tabular}{ccccccccc}
\toprule
Accretion               & Resolution & $\racc$    &   $n_0$    &  $\Delta\Mbh$ & $\Delta M_*$&   $\mSFR$ &     $\Mgas$ \\
--                      &     (pc)   &       (pc) &  ($\cc$)   & ($10^6~\Msun$)&($10^6~\Msun$)&($\Msunyr$)& ($10^{10}~\Msun$)\\
  (1)                   &      (2)   &        (3) &    (4)     &   (5)         &     (6)     & (7)       &  (8)         \\
\midrule
Mass weighted\\
\midrule
BE300\_C\_H             &    300     &    300     &      1     &         60.20 &     2088.61 &     41.77 &     29.35   \\
BE30\_C\_H              &     30     &     30     &      1     &         56.96 &     5702.33 &    114.04 &     29.26   \\
BE3\_C\_H               &      3     &      3     &      1     &         56.65 &     7330.36 &    146.54 &     29.79   \\
\midrule                                                                                                             
BE300\_C\_L             &    300     &    300     &      0.01  &         13.87 &        4.11 &      0.08 &      0.36   \\
BE30\_C\_L              &     30     &     30     &      0.01  &         43.48 &       43.96 &      0.87 &      0.35   \\
BE3\_C\_L               &      3     &      3     &      0.01  &         44.34 &       44.25 &      0.88 &      0.35   \\
\midrule                                                                                                             
Volume weighted\\
\midrule                                                                                                             
BEV300\_C\_L            &    300     &    300     &      0.01  &       7.22    &      73.47  &     1.47  &       0.30   \\
 BEV30\_C\_L            &     30     &     30     &      0.01  &      49.43    &     179.47  &     3.79  &       0.35   \\
  BEV3\_C\_L            &      3     &      3     &      0.01  &     234.47    &       2.40  &     0.05  &       0.35   \\
\midrule
BEV300\_C\_H            &    300     &    300     &      1     &      60.33    &    2268.57  &    45.37  &      29.33   \\
 BEV30\_C\_H            &     30     &     30     &      1     &      57.85    &   21600.51  &   432.01  &      29.18   \\
  BEV3\_C\_H            &      3     &      3     &      1     &      61.06    &    5731.05  &   114.62  &      29.04   \\
\bottomrule
\end{tabular}%
}%
\flushleft
\parbox{0.9\linewidth}{\footnotesize 
\textit{Notes.} Same as Table~\ref{tab1}.}
\end{table*}

\begin{table*}
\caption{Volume weighted simulations.}\label{tab3}
\makebox[\linewidth]{%
\begin{tabular}{ccccccccc}
\toprule
Accretion               & Resolution & $\racc$    &   $n_0$    &  $\Delta\Mbh$ & $\Delta M_*$&   $\mSFR$ &     $\Mgas$ \\
--                      &     (pc)   &       (pc) &  ($\cc$)   & ($10^6~\Msun$)&($10^6~\Msun$)&($\Msunyr$)& ($10^{10}~\Msun$)\\
  (1)                   &      (2)   &        (3) &    (4)     &   (5)         &     (6)     & (7)       &  (8)         \\
\midrule
BV300\_S\_H             &      0.1   &    300     &      1     &          0.17 &    2584.31  &     52.49 &      36.20 \\
BV30\_S\_H              &      0.1   &     30     &      1     &          0.25 &    2715.80  &     55.10 &      36.54 \\
BV3\_S\_H               &      0.1   &      3     &      1     &          0.58 &    1405.10  &     28.52 &      35.19 \\
BV300\_S\_L             &      0.1   &    300     &      0.01  &          0.02 &      29.81  &      0.90 &       5.38 \\
BV30\_S\_L              &      0.1   &     30     &      0.01  &          0.26 &      28.75  &      0.63 &       5.39 \\
BV3\_S\_L               &      0.1   &      3     &      0.01  &          0.47 &      27.95  &      0.56 &       5.38 \\
\midrule                                                                                                             
BV300\_C\_H             &      300   &    300     &      1     &          1.37 &    7489.82  &    149.80 &      28.73 \\
BV30\_C\_H              &       30   &     30     &      1     &         36.10 &    4341.17  &     86.82 &      30.34 \\
BV3\_C\_H               &        3   &      3     &      1     &          0.64 &    2493.44  &     49.87 &      31.73 \\
BV300\_C\_L             &      300   &    300     &      0.01  &          0.31 &      39.46  &      0.79 &       0.35 \\
BV30\_C\_L              &       30   &     30     &      0.01  &          0.51 &      24.67  &      0.49 &       0.36 \\
BV3\_C\_L               &        3   &      3     &      0.01  &          0.88 &       9.53  &      0.19 &       0.37 \\
\midrule                                                                                                             
\midrule                                                                                                             
BEV300\_C\_H            &      300   &    300     &      1     &          1.37 &    7587.78  &    151.75 &      28.70 \\
BEV30\_C\_H             &       30   &     30     &      1     &          9.17 &    5893.80  &    117.88 &      29.70 \\
BEV3\_C\_H              &        3   &      3     &      1     &          0.63 &    2502.28  &     50.84 &      31.91 \\
BEV300\_C\_L            &      300   &    300     &      0.01  &          0.31 &      39.46  &      0.79 &       0.35 \\
BEV30\_C\_L             &       30   &     30     &      0.01  &          0.57 &      29.46  &      0.59 &       0.36 \\
BEV3\_C\_L              &        3   &      3     &      0.01  &          0.37 &       9.01  &      0.18 &       0.37 \\
\bottomrule
\end{tabular}%
}%
\flushleft
\parbox{0.9\linewidth}{\footnotesize 
\textit{Notes.} Same as Table~\ref{tab1}.}
\end{table*}

\subsection{The initial conditions}\label{initial_conditions}
The total gravitational potential is an isothermal sphere with a velocity dispersion $\sigma=150~\kms$, accounting for the stellar and dark matter distributions, plus a point mass potential for the central BH. The gas is not self gravitating.

The stellar distribution is described by a \citet{jaffe1983} profile with a half-light radius of $5.3~\kpc$, a total mass $\Mstar=3\e{10}~\Msun$, and the mass-to-light ratio is assumed to be 5.8 in solar unit in the $B$-band, independently of the position (the structural and dynamical
properties of these galaxy models are discussed in detail in \citealt{ciotti.etal2009b}). The initial BH mass is $\Mbh=10^{-3}\Mstar$, thus placing the galaxy near the Magorrian relation \citep{magorrian1998, kormendy.ho2013}.

The initial ISM density is a spherical distribution described by
\[
 \rho (r) = \dfrac{\mu \mp n_0}{(1+\tilde{r}^2)^{\beta}}, \qquad \tilde{r} \equiv r/\rc,
\]
where $\mp$, $\mu$, $\rc$, $n_0$, $\beta$ are the proton mass, mean molecular weight, core radius, central number density and slope of the ISM distribution, respectively. We adopt $\mu=0.62$ (solar metallicity), for simplicity $\rc=5.3~\kpc$ (consistent with the stellar distribution) and $\beta=1.3$ to obtain a density similar to the critical baryon density in the last radial grid when $n_0=0.01~\cc$ is adopted (see Section~\ref{results}). Then, the temperature profile is calculated by solving the equation of hydrostatic equilibrium for the ISM with a zero pressure boundary condition at the outer edge of the grid.

In principle, in 2D axisymmetric simulations,  gas and stars can have an arbitrary rotational velocity $\uphi$ and $\vphi$ as a function $(r, \theta)$. \citet{negri.etal2014, negri.etal2014b} show that a rotating stellar population is able to drag the ISM, thus producing a gaseous halo almost co-rotating with the stars. However, angular momentum conservation in rotating galaxies would lead to the formation of a cold, rotationally supported disc of kpc-scale \citep{negri.etal2014, negri.etal2014b,  negri.etal2015}, thus preventing any substantial accretion on the central BH.
In real objects, these massive discs would be prone to gravitational instabilities that can transport angular momentum outwards and so drive gas on the central object \citep{bertin.lodato.2001,hopkins.quataert2011}. Yet, this kind of instability cannot be captured in 2D axisymmentric simulations without a prescription for angular momentum transport. In order to allow accretion on the BH, in this work we limit ourself in the low rotation regime, by constraining the centrifugal barrier radius to reside inside the innermost radial grid. 
The local specific angular momentum of the circular orbit in the meridional plane is given by
\[
 j(r,\theta) = \sin^2\theta \sqrt{  r^2\sigma^2 + G\Mbhi r}.
\]
Thus, we impose the following rotational field $\uphi$ ($\vphi$) for the gaseous (stellar) component:
\[
 \dfrac{1}{\uphi} = \dfrac{d}{\sigma R} + \dfrac{1}{f\sigma} + \dfrac{R}{j(\rmin, \upi/2)},\label{rotation}
\]
where $R=r\sin\theta$ is the distance from the axis, $\Mbhi$ is the BH mass at the beginning of the simulation, $d=10~\pc$, $f=1$, $j(\rmin, \upi/2)$ is the specific angular momentum of the circular orbit at radius $\rmin$ in the equatorial plane, and $\rmin$ is the radius of the first grid. The resulting profile gives solid body rotation at $R<d$, and a constant angular momentum distribution at large radii, while the $f$ term guarantees that the velocity field never exceeds $\sigma$. As mentioned in Sec.~\ref{sec:grid}, if a gas blob hits the $\theta$ boundary, it remains ``attached'' to it, thus leading to un unphysical BH accretion and star formation density. In order to avoid this numerical issue, we modified the stellar rotation field $\vphi$ near the $\theta$ boundary, by taking it as the maximum between Eq.~\ref{rotation} and $5\psi(r, \theta) j(r,\theta)/R $, where 
\begin{align}
 \psi(r,\theta) \equiv & \dfrac{1}{8} \left[ 2 +  \tanh \dfrac{\thetamax - \theta }{\sigtheta} + \tanh \dfrac{\thetamax + \theta -\upi}{\sigtheta}\right] \\
 & \times \left[ 1 + \tanh \dfrac{\rmax - r}{\sigr} \right] \left[ 1 + \tanh \dfrac{r - \rmiin}{\sigr} \right], \label{eq:psi}
\end{align}
where $\thetamax=0.15$~rad, $\rmiin=70~\pc$ and $\rmax=300~\kpc$. This formulation provides a smooth transition from the biconical region described by Eq.~\ref{eq:psi} and the rest of simulation box, where the rotation law is provided by Eq.~\ref{rotation}.


\begin{figure*}
\includegraphics[keepaspectratio, width=\textwidth]{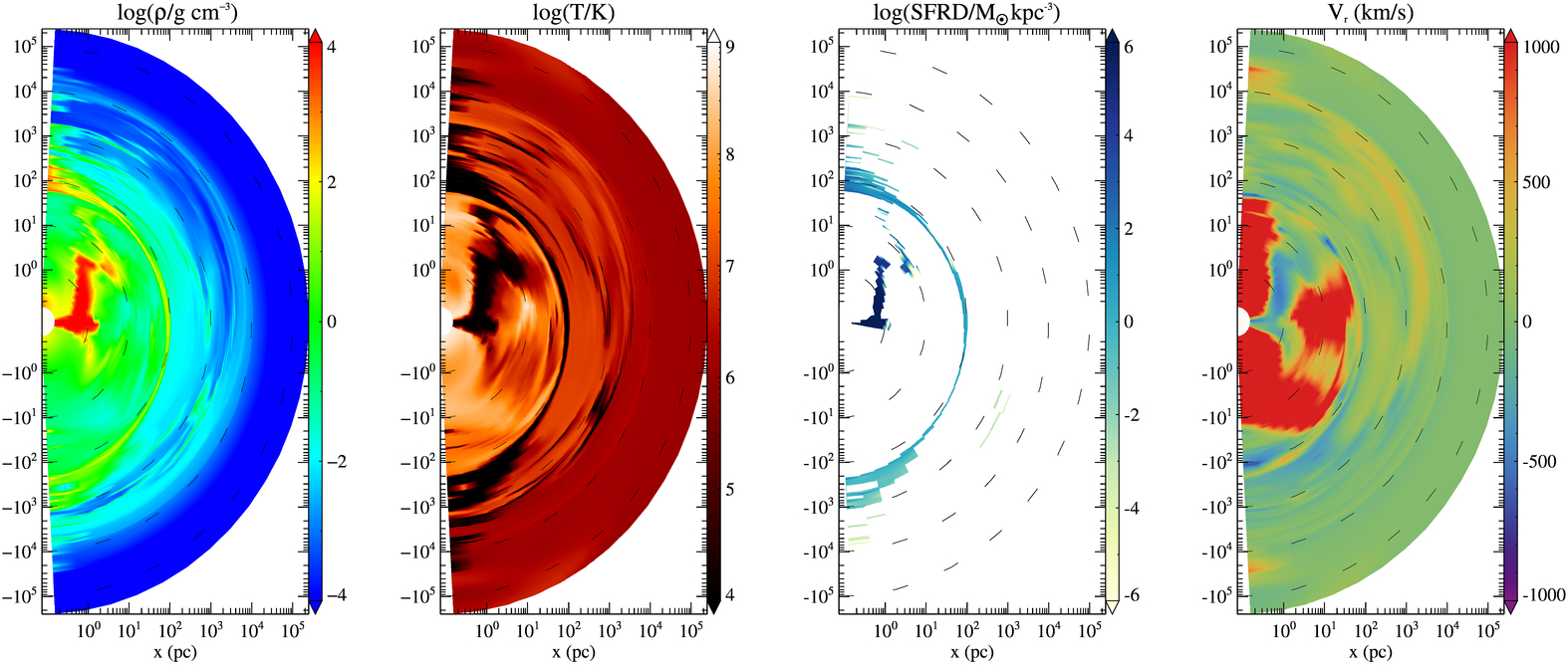}
\caption{From left to right: meridional section of density (in $\cc$, left panel), temperature (in K, middle panel), SFR density (in $\Msun\yr^{-1}\kpc^{-3}$) and radial velocity (in $\kms$, right panel) for the flux accretion simulation with $n_0=0.01~\cc$ at $t=38.7~\Myr$. The figure presents a representative accretion event: the cold, infalling gas is feeding the BH, while the BAL wind begins to inject mass, energy and momentum in the polar direction.}\label{map1}
\end{figure*}
\begin{figure*}
\includegraphics[keepaspectratio, width=\textwidth]{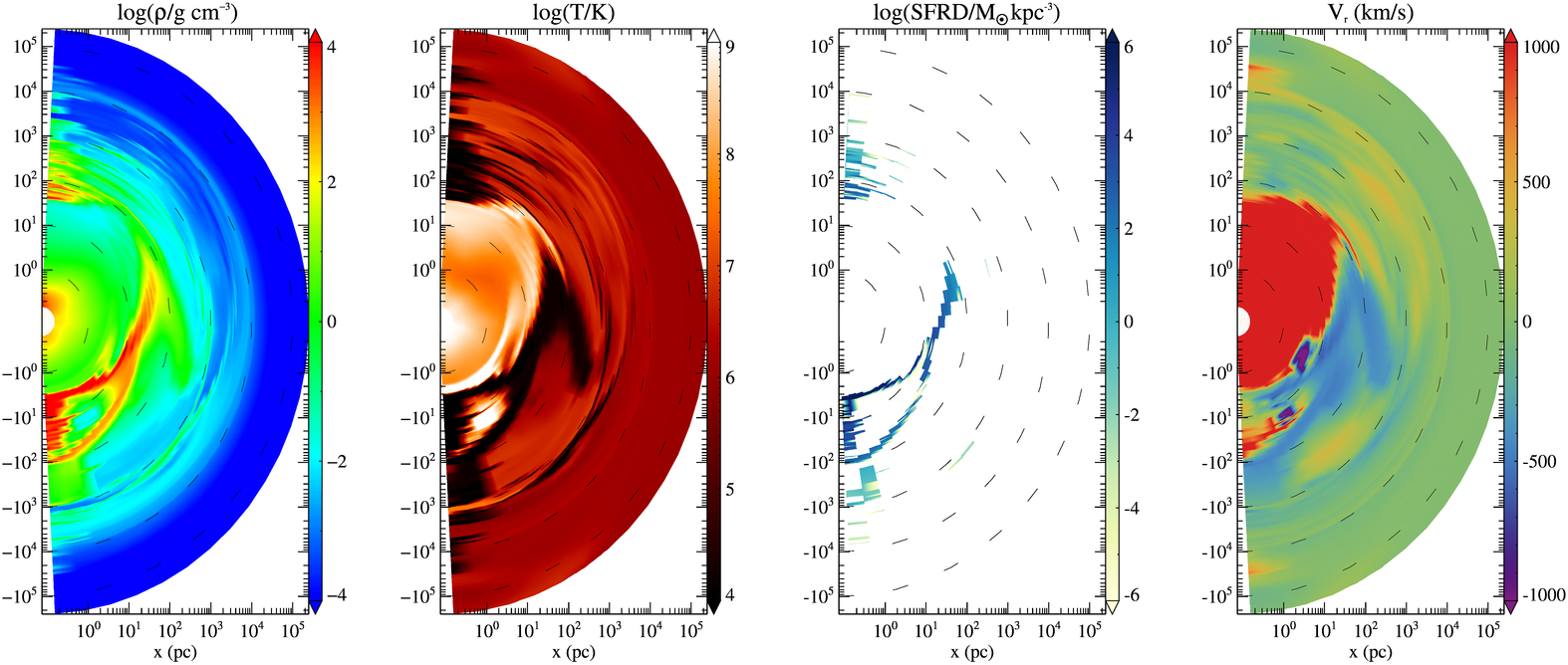}
\caption{Same quantities as Fig.~\ref{map1}, but at the end of the accretion event, at $t=48.7~\Myr$. The BAL wind created a central hot bubble, temporarily shutting down any accretion on the BH.}\label{map2}
\end{figure*}

\section{Results}\label{results}
Here we present the main results of the simulations, summarized in Tables~\ref{tab1}-\ref{tab3}. As reference runs, in Section~\ref{sec:flux} we consider two different runs  employing the flux accretion scheme and 0.1~pc grid resolution, but having different initial gas density ($n_0=$1;~0.01~$\cc$). For each flux simulation we produced two different families with a sub-grid  Bondi scheme: 1) a pure Bondi accretion, as described in Section~\ref{bh_accretion}, and 2) a Bondi accretion capped at the Eddington limit. In each of these families we keep the 0.1~pc hydrodynamical grid, but consider three different values of the accretion radius ($\racc=$3, 30, 300~pc). We retain the 0.1~pc resolution hydrodynamical grid of the flux simulations in order to isolate the effects of the Bondi accretion scheme (Section~\ref{sec:bondi_high_res}) from grid effects. In addition, the mechanical feedback is deposited in the innermost radial grids as in the flux runs, so the ratio of mass/momentum/energy injection and the injection radius is independent of $\racc$. This setup is similar to the BH accretion implementation of \citet{curtis.etal2015}; we discuss few tailored simulations employing their accretion scheme (see Section~\ref{sec:discussion}). Bondi high resolution simulations with density and sound speed calculated as volume weighted quantities are presented in Section~\ref{sec:volume}. Finally, we employ two additional families of runs, with identical initial conditions and BH accretion schemes, but with a central spatial resolution equal to the accretion radius $\racc$ explored in the above families (Section~\ref{sec:bondi_low_res}). 

\AN{Each simulation is run for 100 Myr; the evolution of the main quantities is rescaled in time in the figures, so that the first burst of BH accretion (at $10^{-2}$ the BH Eddington ratio) is at 2 Myr, ending with a presented evolution of 50 Myr. This timescale is longer than the Salpeter time ($\simeq 45~\Myr$) and the crossing time inside the effective radius ($\simeq 35~\Myr$, upper limit in absence of AGN feedback). The simulations do not show a significant change in the trends after this time. We avoid running the isolated galaxies for a much longer timescale because isolated galaxies lack the cosmic context that modulates the gas content and galaxy structure, via gas accretion from filaments and mergers.}

\subsection{Flux accretion simulations}\label{sec:flux}
We present here our two reference simulations, characterized by a high resolution grid and non-parametrized accretion. Despite the different initial gas fractions, the hydrodynamical evolution of both models follows a common picture.

\begin{figure*}
 \includegraphics[keepaspectratio, width=\linewidth]{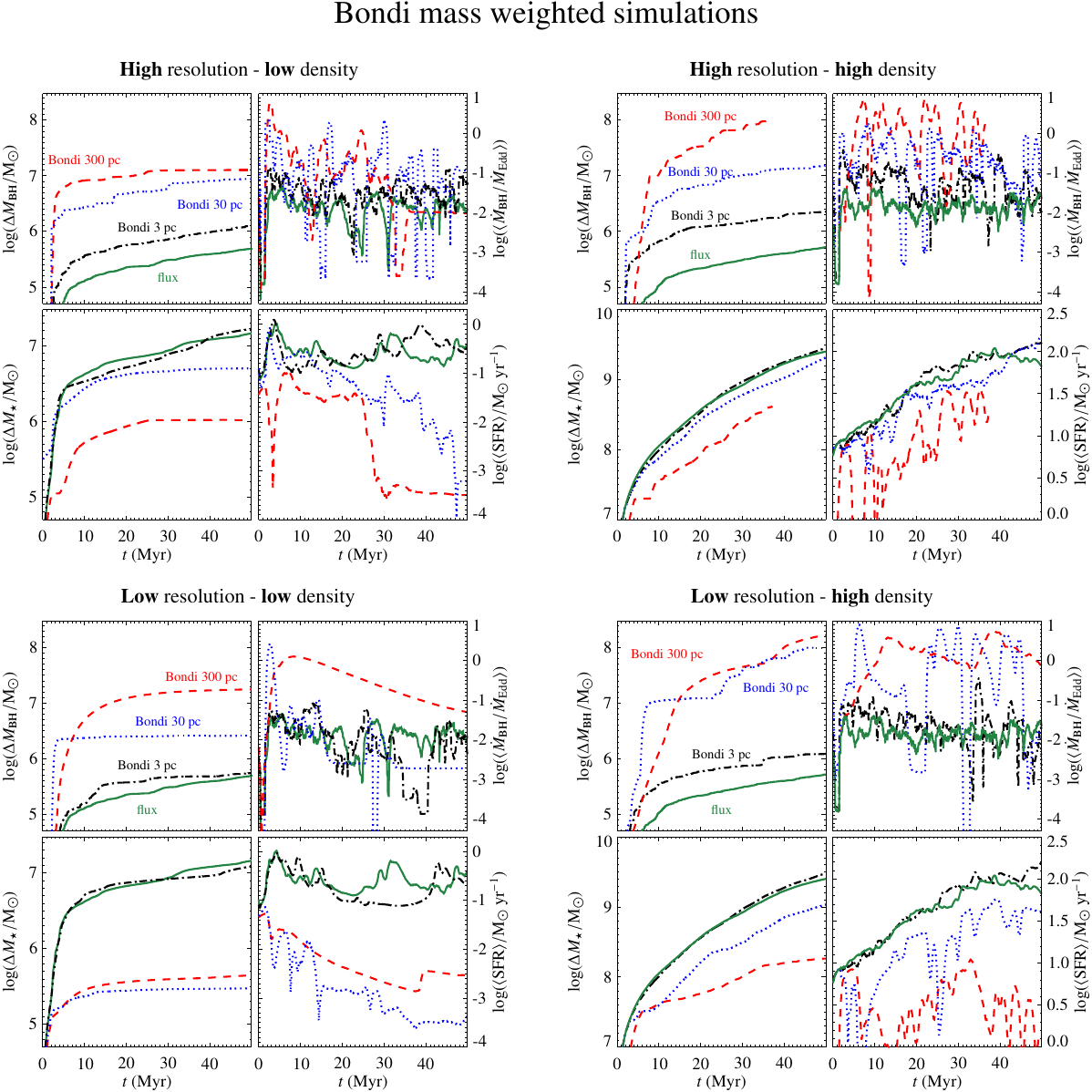}
 \caption{Flux and Bondi mass weighted models evolution with time. Each panel is divided in four quadrants showing BH mass (top left), BH accretion rate (top right), stellar mass formed (bottom left) and SFR (bottom right) as a function of time.  BH accretion rate and SFR are shown as a moving average over $1~\Myr$. The top (bottom) row shows high (low) resolution simulations, while low (high) gas density simulations are presented in the left (right) column.
 The 300 pc run in the top-left panel stopped due to the timestep becoming too short. The flux accretion run never exceeds the Eddington limit. To exclude the formation time of the gas cold shell, all the plots have been shifted in time so that the cold accretion mode (marked when $\Mdotbh/\Mdotedd$ reaches the value of 2 for the first time) takes place at $2~\Myr$.}\label{fig:high_res_bondi}
\end{figure*}

Figure~\ref{map1} shows a typical accretion event: cold, star forming  gas is accreting along the equatorial plane, while the BAL wind begins to inject mass, momentum and energy in the ISM along the  polar directions. Accretion continues until the bipolar outflow creates a hot, elongated expanding bubble of size $\approx 100~\pc$ in the direction of least resistance. BH accretion is thus suppressed (Fig.~\ref{map2}). However,  AGN feedback is not powerful enough to evacuate all the gas present in the galaxy, and after $\sim 10^4-10^5~\yr$ the external cold gas is able to penetrate the hot bubble and to accrete on the BH, thus producing an irregular accretion-feedback-hot bubble cycle that lasts for the entire evolution and self-regulates the BH accretion. Figure~\ref{fig:high_res_bondi} (green line) shows even more clearly the effects of the erratic accretion and AGN feedback cycle: BH accretion rate and SFR time evolution are highly fluctuating (green line, note that this is a moving average over 1~Myr) during the entire evolution. Remarkably, the BH self-consistently limits its own accretion rate to sub-Eddington values via AGN feedback (note that in the flux simulations we do not employ any capping on the BH accretion rate), producing a BH \AN{mass increment} of $\sim 4\e{5}~\Msun$ \AN{(see Table~\ref{tabflux})}.

The top-left panel of Figure~\ref{fig:high_res_bondi} shows the temporal evolution of BH mass, BH accretion rate, stellar mass and star formation rate (SFR) for the flux simulation having $n_0=10^{-2}~\cc$. After a few cooling times, a cold, infalling shell is formed in both simulations at $r\gtrsim 1~\kpc$. The shell quickly fragments due to the Rayleigh–Taylor instability, producing a multiphase ISM characterized by a hot, diffuse component plus clumpy cold gas. This leads to a clumpy, cold accretion that efficiently feeds the BH \citep[for a comparison between 1D and 2D simulations see][]{novak.etal.2011}. 

The counterpart at high gas density (top-right panel in Fig.~\ref{fig:high_res_bondi}) shows a much higher star formation rate, with a final stellar mass about two orders of magnitude higher, but, strikingly, the BH accretion rate and global growth are not  boosted by the higher gas fraction, staying stable at a fraction about few per cent of the Eddington rate. AGN feedback seems to be more effective at regulating BH growth than star formation in the galaxy. 

\subsection{Bondi simulations with a high resolution grid} \label{sec:bondi_high_res}

%
%
%
%
Moving from flux accretion to the Bondi accretion runs, we show here the set of simulations employing the high resolution grid. BH accretion is computed from equation~(\ref{eq:bondi}) having an accretion radius $\racc=3$, 30 and $300~\pc$ (see Section~\ref{bh_accretion}), and the accretion radius is resolved respectively by 23, 98 and 208 grid elements in the radial direction.


The top-left panel in Figure~\ref{fig:high_res_bondi} presents the time evolution of the Bondi models having $n_0=10^{-2}~\cc$. The most striking feature is the apparent trend of $\Delta\Mbh$ with $\racc$, in contrast with the usual assumptions in the literature that an unboosted Bondi formalism underestimates the accretion rate. The larger the accretion radius, the more the BH mass accretion is overestimated with respect to  flux accretion  (see also Table~\ref{tab1}). 

A common feature of all the runs employing the high resolution grid is the concurrent formation of the cold shell at the same time of the flux simulation. The BH in the 300 pc run (red line in the top-left panel of Fig.~\ref{fig:high_res_bondi}) begins to accrete when the cold shell falls inside the accretion radius, reaching immediately a state of super-Eddington accretion. This produces a strong BAL wind that reduces the amount of cold gas inside $\racc$ to $6\e{6}~\Msun$. Following the same accretion-feedback-hot bubble cycle of the flux simulation, a hot expanding bubble is formed, and star formation is quenched. This cycle holds until the BAL wind is finally able to evacuate all the cold gas from the innermost $10~\kpc$ at $t=32~\Myr$. From then onwards, no cold gas is formed and the BH is in a hot, sub-Eddington accretion mode. 

The 30 pc run shows a similar evolution of $\Delta\Mbh$ with respect to the 300 pc simulation; $\Mdotbh$ fluctuates from super-Eddington accretion to strongly sub-Eddington, while $\Mdotbh$ in the 3 pc case is oscillating at higher frequency, and accretion remains overall lower than in the two previous cases (the Bondi 3~pc model accretes at super-Eddington rate at $t\simeq 2~\Myr$, but this phase is not visible in figure due to the moving average).

AGN feedback is much more effective for smaller $\racc$ since the volume to be cleared from cold material is smaller. The amount of cold gas inside $\racc$ for the 30 and 3 pc runs is smaller than the 300 pc simulation, having less than $10^4$ and $10~\Msun$ of cold gas inside $\racc$ for the 30 and 3~pc model, respectively (note that in the last case the BH is mainly in a state of hot accretion). In the 300~pc case, accretion proceeds until $\Mdotbh$ becomes so large that feedback sweeps away all the cold material in one very strong event, producing a fast wind and fierce heating. Instead, in the 30 and 3~pc runs  feedback is more effective in removing the cold gas at $r<\racc$ at early times; hence $\Mdotbh$ is prevented from growing to high values.
\begin{figure}
\centering
 \includegraphics[keepaspectratio, width=0.7\linewidth]{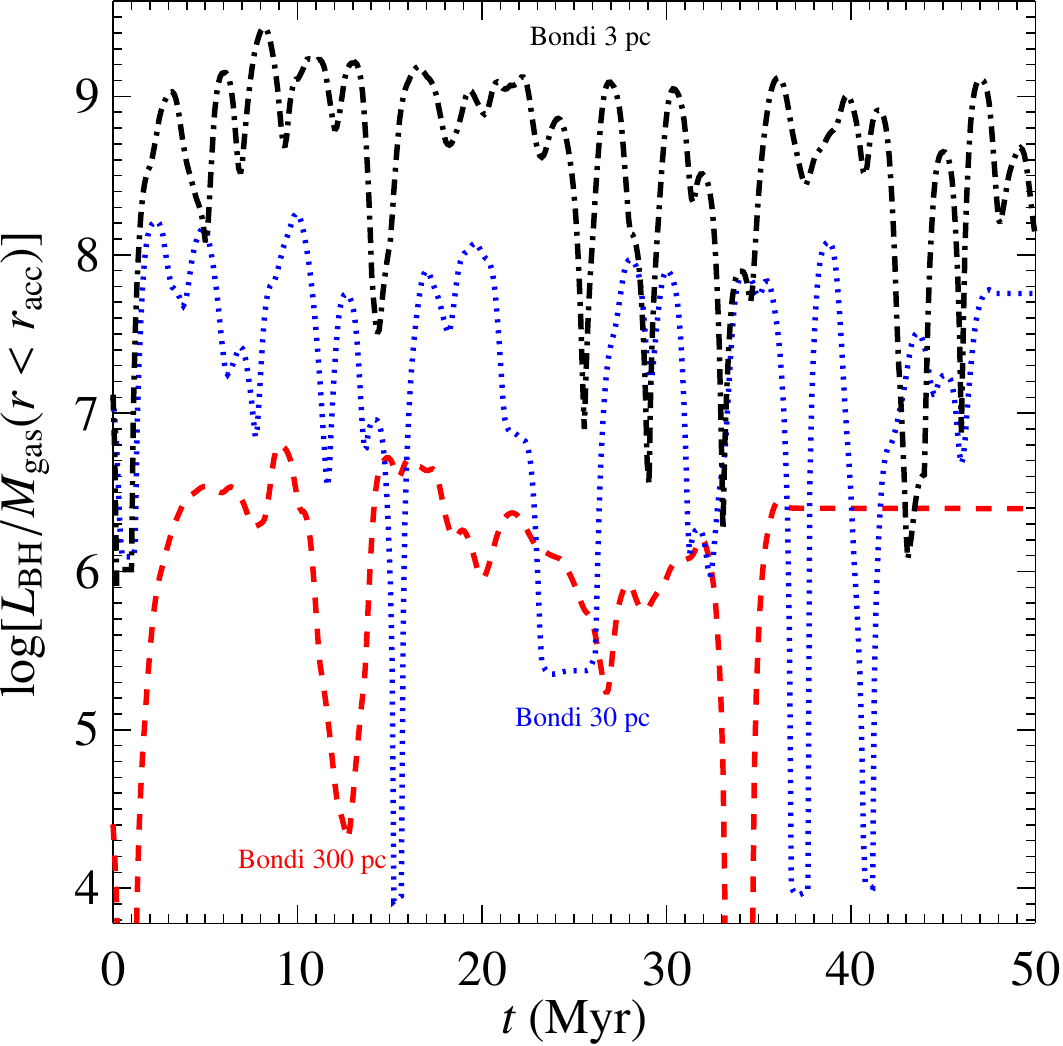}
 \caption{Time evolution of the BH luminosity over gas mass inside the accretion radius (in cgs units) for high resolution, low density simulations (top left panel of Fig.~\ref{fig:high_res_bondi}).} \label{fig:Lbhovermass}
\end{figure}

This is the physical reason of the unexpected positive correlation between accreted mass and accretion radius: with large accretion radii the AGN feedback has to sweep away or heat a larger amount of gas to prevent accretion, thus, the feedback efficiency depends on $\racc$. In Fig.~\ref{fig:Lbhovermass} we quantify the efficiency with the ratio between $\Lbh$, which regulates AGN feedback, and the gas mass inside $\racc$, which has to be displaced or heated to stop the accretion. A clear trend with $\racc$ is present: even if the 3~pc simulation has the least luminous AGN (and thus the weaker feedback), it shows the highest ratio of $\Lbh/\Mgas$. AGN feedback is more effective in clearing the accretion sphere for simulations with small $\racc$.

The trend of BH accreted mass and AGN feedback strength with $\racc$ \AN{at fixed resolution} affects also the amount of gas mass converted into stars. On the one hand in simulations with small $\racc$  feedback limits BH accretion, therefore more cold material outside $\racc$ is left to fuel star formation. On the other hand, for the largest $\racc$ (the 300 pc run), even the cold gas at large radii is heated at $t\simeq 3~\Myr$ by the very strong AGN feedback, prompting a global reduction of cold gas mass and star formation.


To complete the set of high resolution simulations, we present in Figure~\ref{fig:high_res_bondi} (top-right panel) the Bondi runs having $n_0=1~\cc$. They resemble their low density counterparts, in particular the BH mass increases with increasing $\racc$. The physical mechanism is the same, at large $\racc$ the AGN is less efficient in preventing cold gas accretion, producing BH masses up to 100 times the final mass of the flux BH simulation. \AN{The stellar mass shows the opposite trend: the larger $\racc$, the smaller $\Delta \Mstar$, due to larger AGN luminosities at larger $\racc$. The effects of AGN feedback on SFR are minimized in the high density case, because star formation occurs mainly in the massive outer cold shell, at kpc-scales, where gas is sufficiently far that the heating of the AGN feedback is unimportant, but the density is high enough to form stars.}  The introduction of the Eddington limit does not affect the SFR, while the BH final mass is reduced up to a factor of $\simeq 90$ for the 300~pc case (see Tab.~\ref{tab1}).


\subsection{Bondi low resolution grid simulations}
\label{sec:bondi_low_res}

In this section we present the simulation families employing progressively low resolution hydrodynamical grids, with central resolution of 3, 30 and 300~pc. For each grid, $\racc$ is resolved by one cell (in the radial direction), therefore the mechanical feedback is injected in an area that grows with $\racc$.
\subsubsection{Simulations without AGN feedback}
\label{sec:bondi_low_res_noagn}
Given the highly non-linear behaviour of the hydrodynamics, and the complex interplay between accretion and feedback, we first performed a subset of simulations without AGN feedback (see Table~\ref{tab2}) by setting $\epsem$ and ${\epsw}_0 = 0$ in equations~(\ref{eq:epsem}) and (\ref{eq:epsw}).

%
%

\begin{figure*}
\includegraphics[keepaspectratio, width=0.7\linewidth]{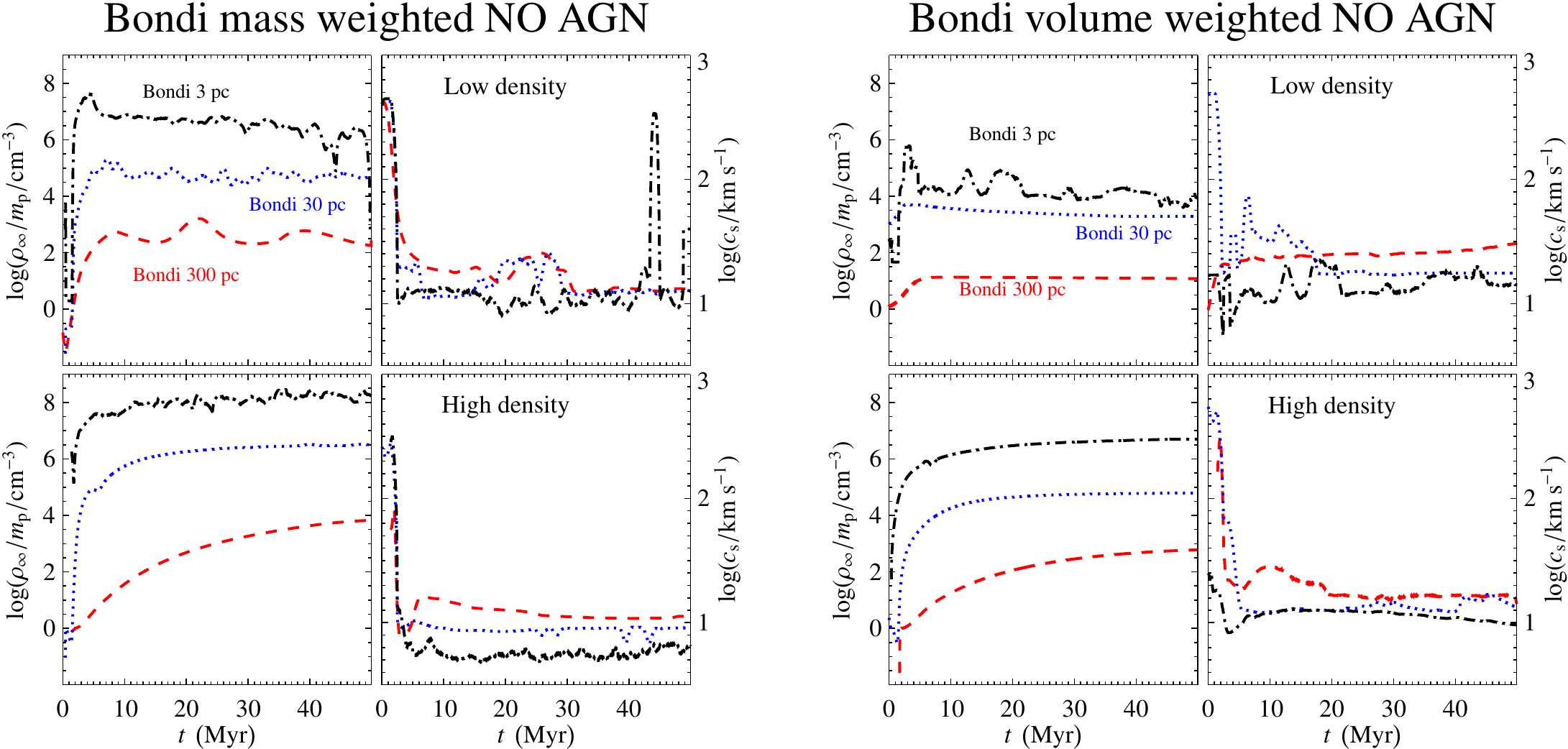}
 \caption{Mass weighted density and sound speed inside $\racc$ for 
 all the models without AGN feedback (see Tab.~\ref{tab2}). \AN{Top and bottom panels refer to low and high gas density simulations, respectively.}}
 \label{fig:rhocs_noagn}
\end{figure*}

\begin{figure}
\centering
\includegraphics[keepaspectratio, width=0.8\linewidth]{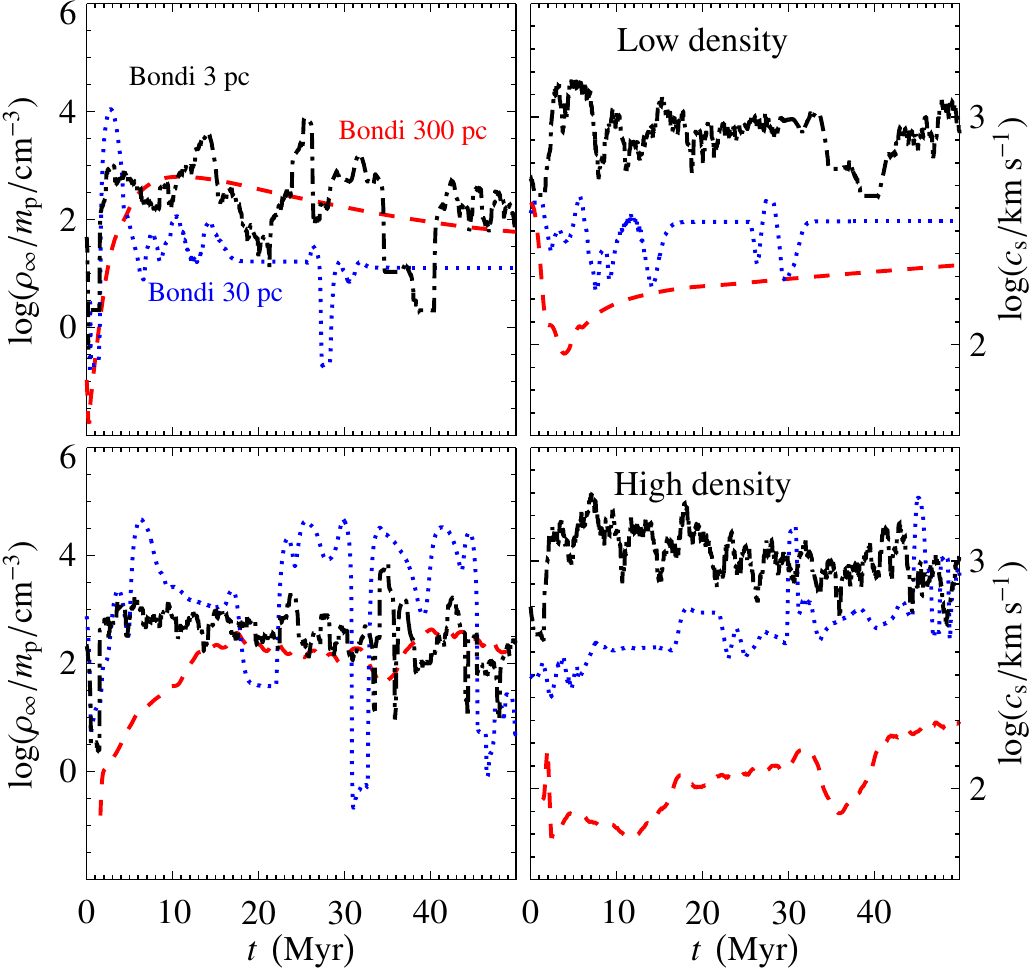}
 \caption{Mass weighted density and sound speed inside $\racc$ for 
 the B3\cl, B30\cl and B300\cl model (top row), and B3\ch, B30\ch and B300\ch model (bottom row). The corresponding simulations employing Eddington limited accretion present the same evolution.} \label{fig:rhocs_agn}
\end{figure}

%

As in the high resolution runs, a cold shell is formed concurrently in all the models due to the cooling of the hot halo. However, the cold shells fragment in BE3\cl and BE30\cl, producing cold clumps that have an easier way to the BH than the main shell body. Instead, in the BE300\cl case the shell proceeds towards the centre in its integrity due to the lower resolution, delaying accretion with respect to the former model.

The lack of feedback has a major impact on the global evolution of all the systems. As soon as the cold shell falls inside the accretion radius, a cooling flow is established until all the cold material present in the galaxy has been accreted. The abundance of cold gas and the absence of a powerful heating mechanism prevent the formation of the accretion-feedback-hot bubble cycle, producing a smooth evolution of BH accretion rate and a steeper stellar density profile in the central 100~pc.

Figure~\ref{fig:rhocs_noagn} presents the time evolution of $\rho_{\infty}$ and ${\cs}_{\infty}$ at $r<\racc$,  the trend usually assumed when $\alpha$ boosting is applied: the higher the resolution, the denser and colder the gas. This can be explained by the fact that at $r\gtrsim\racc$ the gas is better resolved in high resolution runs. As a consequence, smaller and smaller clumps of gas can be resolved during the infall, thus producing denser and colder clouds that prompt higher values of $\rho_{\infty}$ when they cross $\racc$ and ending up with higher Bondi accretion rates.
\subsubsection{Simulations with AGN feedback}
The introduction of AGN feedback entirely modifies the scenario of the previous Section. In Fig.~\ref{fig:rhocs_agn} we can contrast the (mass weighted) density and sound speed that enter the Bondi formalism with and without AGN feedback. The  mean gas temperature inside the accretion radius is higher for smaller resolutions, while the gas densities are all very similar, after the cold gas shell has reached $\racc$. In the 3~pc resolution run the density is much lower than in the case without AGN feedback, while in the 300~pc runs the densities are comparable.
\AN{The physical reason can be again identified in the dependence of the AGN relative strength on the injection radius: the smaller the volume into which AGN feedback is injected, the more efficient it becomes,} i.e. the energy input per unit gas mass in the accretion region is higher the smaller is $\racc$ (cf. Fig.~\ref{fig:Lbhovermass}). We can conclude that the common assumption that low resolution grids lead to an underestimate of $\Mdotbh$ is not verified in presence of resolved and realistic AGN feedback.

The BH and stellar evolution of the low density-low resolution runs are presented in the bottom-left panel of Fig.~\ref{fig:high_res_bondi}, showing the same trends of the high resolution runs, due to the analogous physical evolution of the gaseous phase: the cold shell is formed, it accretes on the BH\footnote{Note that the B300\cl shows a delayed accretion like its counterpart without AGN feedback due to resolution effects, see Section~\ref{sec:bondi_low_res_noagn}.} and the accretion-feedback-hot bubble cycle typical of the high resolution runs is established. The 3~pc resolution run shows high frequency oscillations in $\Mdotbh$, where the 30 and 300~pc runs present a much smoother accretion history. In all cases AGN feedback is able to halt the accretion on short timescales.

%
%
%

The trend of $\Delta\Mbh$ increasing with the size of $\racc$ is in contrast with respect to the no AGN simulations, while it matches the scenario of high resolution runs (Fig.~\ref{fig:high_res_bondi}), where $\Delta\Mbh$ \textit{increases} with the accretion radius.  The interplay between BH feeding and AGN feedback  at different resolutions leads to an accreted BH mass overestimated up to $\sim$ 30 times with respect to the flux simulation, while stellar mass can be underestimated by a factor of $\sim$ 15.

The same trend of $\Delta\Mbh$ increasing at decreasing resolution appears in the high density-low resolution runs shown in the bottom-right panel of Fig.~\ref{fig:high_res_bondi} and in the bottom panel of Fig.~\ref{fig:rhocs_agn}. Here, the mass accretion becomes largely super-Eddington in the 30 and 300~pc runs due to the large gas mass present in the accretion radius. On the other hand, the cold shell, while falling towards the BH, is heated by the AGN radiation. In the 3~pc runs (both at low and high density), the accretion is mainly sub-Eddington, but at larger $\racc$, $\Lbh$ is high enough to heat the cold gas still outside the accretion radius, thus keeping low the SFR.

With degrading grid resolution, at early times, before the BH start accreting and AGN feedback to operate, star formation is higher in the better resolved grid, as densities and temperatures are respectively higher and lower. As soon as AGN feedback is unleashed, star formation gets highly suppressed.


\subsection{Volume weighted Bondi runs}\label{sec:volume}
\begin{figure*}
\includegraphics[keepaspectratio, width=\linewidth]{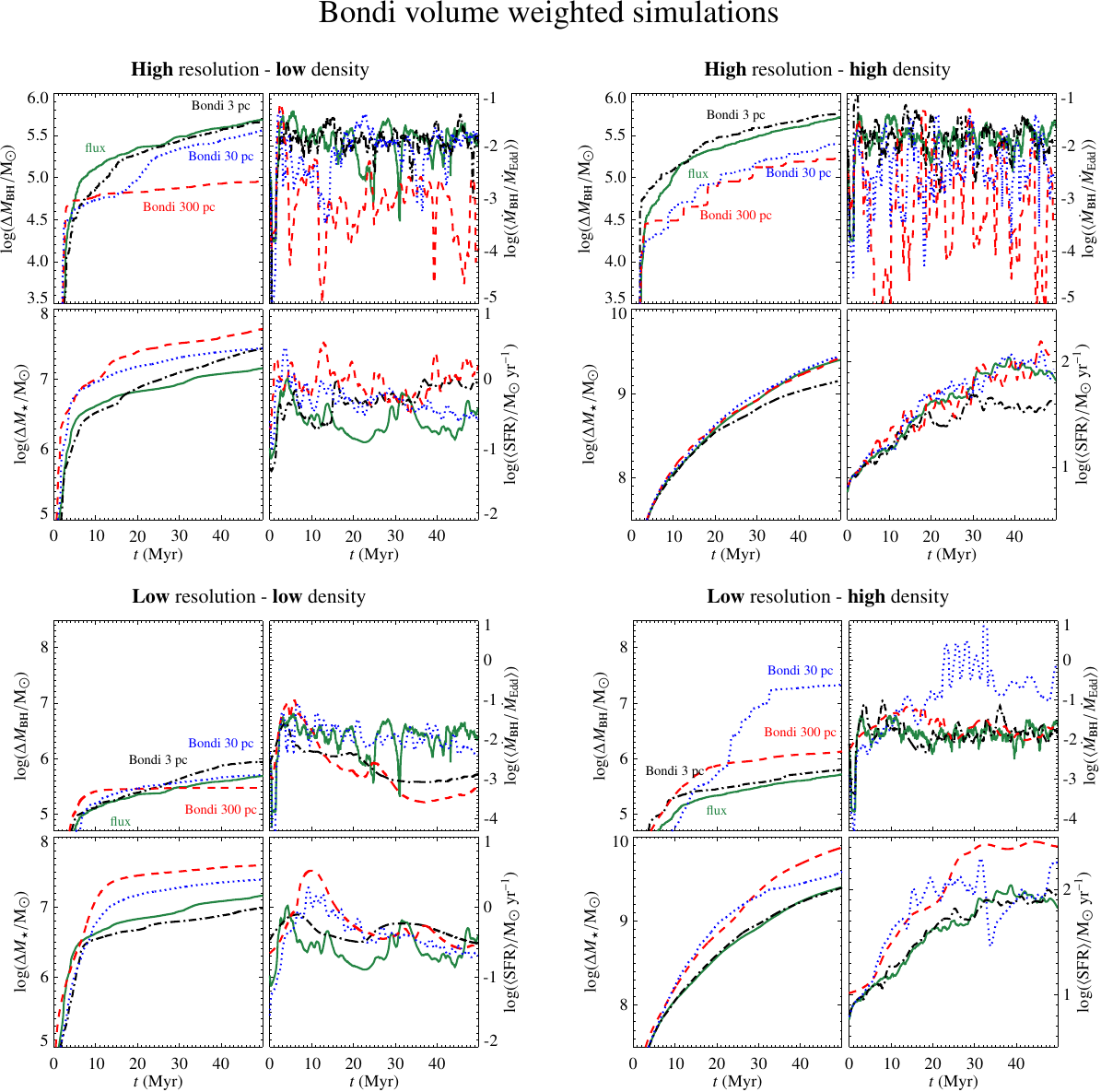}
 \caption{Same as Figure~\ref{fig:high_res_bondi}, for the Bondi $n_0=10^{-2}$ and $1~\cc$ simulations employing volume weighted quantities in equation~(\ref{eq:bondi}) (left and right figure, respectively).} \label{fig:volume_low}
\end{figure*}

An alternative to mass weighting to compute $\rho_\infty$ and ${\cs}_\infty$ is to use volume weighted quantities \citep[e.g.,][and SPH codes employing kernel weighted quantities]{vogelsberger.etal2013, vogelsberger.etal2014}. We present here a family of high resolution simulations employing a volume weighted average of $\rho_\infty$ and ${\cs}_\infty$ in equation~(\ref{eq:bondi}). All the remaining parameters (e.g. the accretion radius $\racc$), initial conditions and grids are unchanged from Section~\ref{sec:bondi_high_res}.

Figure~\ref{fig:volume_low} show the BH and stellar mass evolution, compared with the corresponding flux simulations, for both the high and low density runs. In both cases, at variance with the mass weighted simulations of Fig.~\ref{fig:high_res_bondi} (top panels), the accreted mass is smaller for large $\racc$, since the computed Bondi accretion rate is sensitive to volume effects through $\rho_\infty$ and ${\cs}_\infty$, while at small radii the simulations converge to the flux runs. This can be explained by the hot gas naturally having a filling factor larger than the cold gas inside the accretion radius, therefore the hot phase is the main actor in volume weighted quantities. As a consequence, in the BV300\_S\_L (where the hot gas filling factor is larger and thus its influence is maximised) the BH is in a state of hot accretion for most of the evolution, until a significant volume of cold gas crosses $\racc$, prompting accretion (still sub-Eddington, see $t\simeq 2~\Myr$ in Fig.~\ref{fig:volume_low}, left panels); the subsequent feedback then removes the remaining cold gas and restores the hot accretion mode. As expected, in the high density counterpart the deviation from the flux accretion at large $\racc$ is minimized by the large amount of gas that can cools down, thus enhancing the accretion rate by increasing the cold gas filling factor.

SFR and $\Delta\Mstar$ in low density cases show a weaker dependence on $\racc$ compared to the mass weighted simulations (cf top rows of Figs.~\ref{fig:high_res_bondi} and \ref{fig:volume_low}), since the main situ of star formation is localized in the cold shell, which is formed at $r\gtrsim 1~\kpc$. Moreover, all the simulations spend almost all their evolution at low accretion rates (with mean values of $\Mdotbh/\Mdotedd$ ranging from $4\e{-4}$ to $10^{-2}$), thus the corresponding AGN feedback is unable to significantly affect star formation. The similarities between runs with different $\racc$ are even less noticeable in the high gas density case (Fig.~\ref{fig:volume_low}, right panels), due to the more massive cold shell, where the majority of star formation takes place.

The scenario is modified when the grid is coarser. 
On the one hand, in low resolution runs without AGN feedback the evolution of gas density and temperature inside the accretion radius shows the same trend of mass weighted simulations (cf. left and right panels of Fig.~\ref{fig:rhocs_noagn}), with lower density values attributable to volume weighted mean.
On the other hand, when  AGN feedback is present, the models with $\racc=3$ present a global evolution of $\Delta\Mbh$ and $\Delta\Mstar$  similar to the flux accretion runs. Instead, at larger $\racc=300$ feedback is less efficient in stopping the accretion, producing a larger accretion rate than in the high resolution runs. 
This phenomenon compensates the reduced accretion rate typical of volume weighted Bondi accretion, ending in a BH mass of the same order of magnitude of the flux run for the $\racc=300~\pc$, whereas in the mass weighted case it was overestimated by a factor of 100.  The BV30\_C\_H presents a peculiar evolution, characterized by a oscillating BH accretion rate at $t>20~\Myr$. This is due to a large blob of cold gas entering in the accretion radius, reducing $\cs$ and thus increasing the accretion rate.
In addition, the general time evolution of BH and stellar mass is smoother with respect to the low resolution mass weighted models, being dominated by hot gas, and lacking the strong, super-Eddington peaks triggering strong feedback phases.

%

\subsection{Eddington rate limiter?}


In the runs discussed above, the BH can grow theoretically without limitations. However, in cosmological simulations BH accretion is usually capped at the Eddington luminosity, by assuming $\Mdotbh = \min (\Mdotbondi, \Mdotedd)$. Thus, we simulate the same set of galaxies presented in the previous sections by employing Bondi accretion (mass and volume weighted), capped at the Eddington rate, i.e., assuming a maximal accretion rate corresponding to the Eddington luminosity for a radiative efficiency of 10\%. For the majority of the models this additional constraint does not significantly modify the global picture (see Tables~\ref{tab1} and \ref{tab2}-\ref{tab3}). 
The reduced accretion rate produces a less energetic feedback that is not able to evacuate the accretion volume, maintaining the BH accretion rate at the Eddington limit for the entire evolution. Smaller BH masses are thus produced with respect to the Bondi runs, with the solely exception of the BE300\_S\_L model, due to the fact that in its Bondi counterpart spends half of its evolution in a sub-Eddington phase (cf. with the red line in top-left panel of Figure~\ref{fig:high_res_bondi}). Given the low accretion rates involved, the volume weighted simulations capped at the Eddington rate reported in Table~\ref{tab3} do not show any significant difference with respect to the Bondi accretion runs (except for the BV30\_C\_H run).

\section{Discussion and comparison with previous works}\label{sec:discussion}

Our analysis of the Bondi accretion  in high and low resolution simulations has revealed how BH feeding and AGN feedback influence each other, and that the exact implementation affects the results. Usually, cosmological simulations justify the adoption of a boost parameter $\alpha$ to artificially increase Bondi accretion with two main arguments: (1) large scale simulations do not have the necessary resolution nor the physics (e.g., molecular cooling) to resolve the cold gas, thus the gas temperature near the AGN may be overestimated; (2)  the Bondi radius is normally unresolved, therefore the ISM density may be underestimated.
\begin{figure*}
\centering
 \includegraphics[keepaspectratio, width=\linewidth]{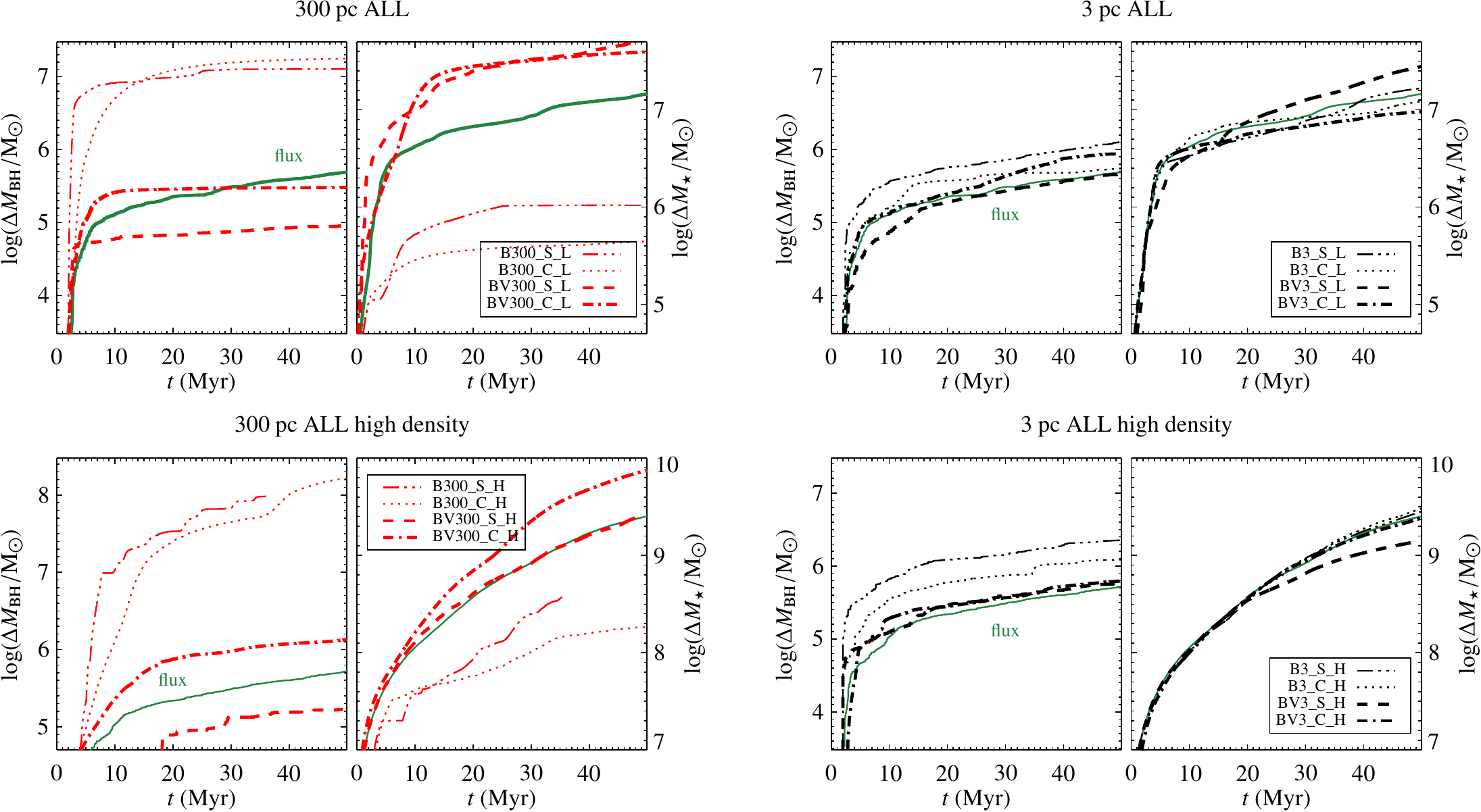}
\caption{Time evolution for the 300~pc (left) and 3~pc Bondi (right) runs having $n_0=10^{-2}~\cc$ (top) or $n_0=1~\cc$ (bottom), compared with flux accretion (simulations capped at the Eddington rate are not included).}
 \label{fig:summary}
 \end{figure*}

We show that the above assumptions, while they remain true in simulations without AGN feedback, are not always verified in its presence. Our reference model employing flux accretion with a 0.1~pc resolution grid display moderately low BH accretion. This is due to feedback, which is able to evacuate the BH surroundings, halt accretion and keep the BH in a self-regulation regime. A cycle of cooling, accretion, feedback and hot bubble is created, that ultimately leads to a cold, irregular accretion. \AN{This accretion-feedback loop is not new in the literature: using 1D simulations \citet{ciotti.ostriker1997, ciotti.ostriker2001, ciotti.ostriker2007} found that the flickering of the AGN is due to cold shells of gas accreting on the BH. Higher frequency oscillations in $\Lbh$ are produced in 2 and 3D simulations, where the gas cold shell naturally fragments, producing a rain of cold gas clumps that can dramatically increase $\Mdotbh$ (this accretion mode has been dubbed Cold Chaotic Accretion in \citealt{gaspari.etal2013, gaspari.etal2016}, see also \citealt{ciotti.etal2016}).}

\AN{We explored the parameter space by varying the accretion algorithm, the accretion radius and the numerical resolution of the simulations. Comparison between different models can be performed by either keeping constant the resolution, and comparing simulations with different $\racc$, or keeping fixed $\racc$ and considering different resolutions, in order to isolate pure resolution effects. Finally, one could consider comparing simulations with different resolutions and $\racc$, but with a specified accretion algorithm.}

\subsection{Comparison at constant resolution with various $\racc$}\label{sec:fix_resolution}
\AN{From Figure~\ref{fig:scatter2}, by keeping fixed the resolution (filled and empty circles), a clear trend emerges: in the case of mass weighted accretion (left panel), \textit{$\Delta\Mbh$ increases and $\Delta\Mstar$ decreases at increasing $\racc$.} When $\racc$ is much larger than the Bondi radius (red filled circles) the BH mass is overestimated up a factor of 100. In volume weighted simulations, instead, the trend is opposite: $\Delta\Mbh$ and $\Delta\Mstar$ are respectively under and over-estimated with respect to the flux case, at increasing $\racc$.}

This difference is due to the complex interplay between accretion and feedback. We show that the feedback efficiency in clearing the accretion region is a function of the accretion radius.
\AN{On the one hand, in the mass weighted case, which is dominated by accretion of cold gas, at large $\racc$  AGN feedback is not effective in evacuating the accretion region, since the mass accumulated is too large to be pushed outside $\racc$ or heated up. This leads to a continuous accretion that proceeds until $\Lbh$ becomes so large that feedback generates a hot bubble that shuts off accretion. However, at this point the BH has already increased its mass by a considerable amount with respect the flux accretion run. In this regime, the BH accretion rate should be de-boosted to recover the evolution of the flux accretion simulation.}

\AN{On the other hand,} the volume average is essentially dominated by the hot and cold gas filling factors and not by their mass. Thus, at large $\racc$, where the filling factor of the hot gas dominates over that of the cold gas, the BH is in a state of hot accretion, leading to a lower BH final mass with respect to the flux case (see empty circles in Fig.~\ref{fig:scatter}). In this regime, that is what is normally assumed in literature, boosting the accretion rate when large $\racc$ are used goes in the right direction.


This picture is strengthened when an Eddington capping is applied to Bondi accretion: in the mass weighted case AGN feedback is further hindered by the capping, leading to continuous accretion for the entire evolution, whereas in the volume weighted case there is no sensitive difference.

\subsection{Comparison at constant $\racc$ with various resolutions}
\begin{figure*}
\centering
 \includegraphics[keepaspectratio, width=0.8\linewidth]{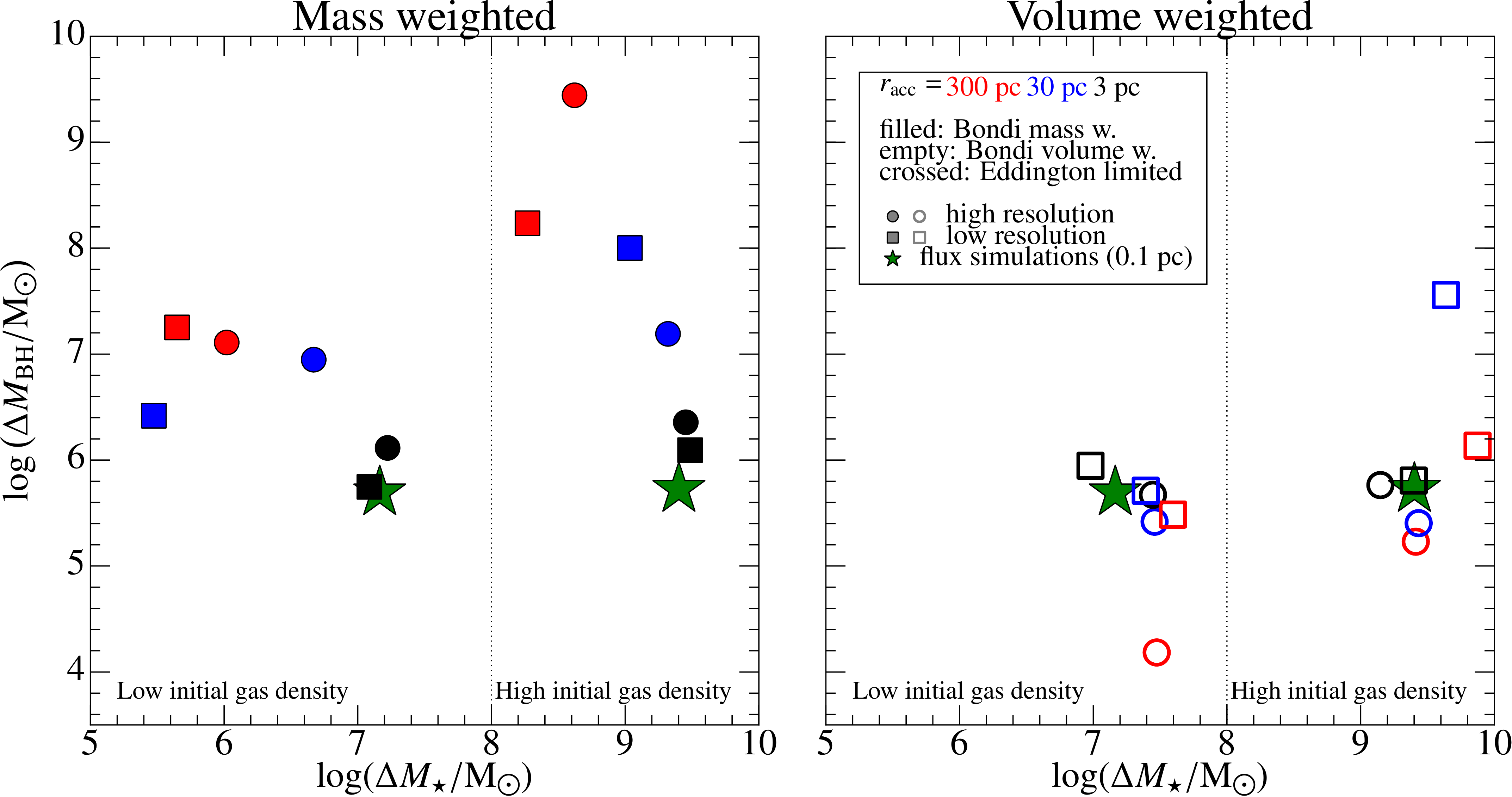}
\caption{Accreted BH mass vs stellar mass for the simulations of Tables~\ref{tab1}-\ref{tabflux} and Table~\ref{tab3}, except Eddington limited runs. Circles and squares represent runs with standard and coarse grid, respectively, while filled and empty symbols refer to mass and volume weighted simulations. Crossed symbols are Eddington limited runs. The accretion radius is mapped by colours: red, blue and black represent $\racc=300$, 30 and 3~pc. Green stars represent the flux simulations (which employ the standard 0.1~pc resolution grid). All the simulations with $\Delta \Mstar<10^8~\Msun$ (vertical dotted line) have $n_0=10^{-2}~\cc$, $n_0=1~\cc$ otherwise. The models converge towards the flux runs when the Bondi radius is progressively better resolved. Mass weighted simulations always overestimate $\Delta\Mbh$, high resolution volume weighted simulations underestimate it, while for low resolution volume runs a clear trend is not present.}
 \label{fig:scatter2}
 \end{figure*}

\AN{In considering the pure effect of resolution, we can compare high and low resolution simulations sharing the same $\racc$ (represented by squares and circles of the same colour in Figure~\ref{fig:scatter2}). In the mass weighted case no clear trend is visible\footnote{Note that the B30\_C\_L simulation shows a much smaller $\Delta\Mstar$ with respect to what the trend of the other models suggest due to a very early, and short-lived phase of strong AGN feedback, suppressing star formation. See Fig.~\ref{fig:high_res_bondi}, bottom-left panel, $t\sim 2$ Myr.}, except at low $\racc$ (black symbols), where $\Delta\Mbh$ is lower at low resolution, although the two simulations are very similar. Surprisingly, in volume weighted runs $\Delta\Mbh$ increases at decreasing resolution with constant $\racc$, showing an evolution similar to the flux case (with the notable exception of the BV30\_C\_H).}

In principle one would expect that the lower the resolution, the worse the ability of the simulation to capture the properties of the gas feeding the BH, located, in reality, on sub-pc scales. This result is caused by a combination of two factors cancelling each other out. On the one hand, volume weighting tends to {\it underestimate} BH accretion, because of the hot gas filling factor dominance, and this behaviour is enhanced at large $\racc$ (see also  Section~\ref{sec:fix_resolution}). On the other hand, at low resolution and large $\racc$, AGN feedback is less effective in clearing out cold gas. The combination of these effects produces reasonable BH and stellar masses, although stellar mass may be overestimated in gas-rich galaxies (see Fig.~\ref{fig:summary}). This surprising compensation may explain why, statistically, BH and galaxy properties are well reproduced in large-scale, relatively low-resolution cosmological simulations \citep[see the discussion in][]{volonteri.etal2016}.


 \begin{figure}
\centering
 \includegraphics[keepaspectratio, width=\linewidth]{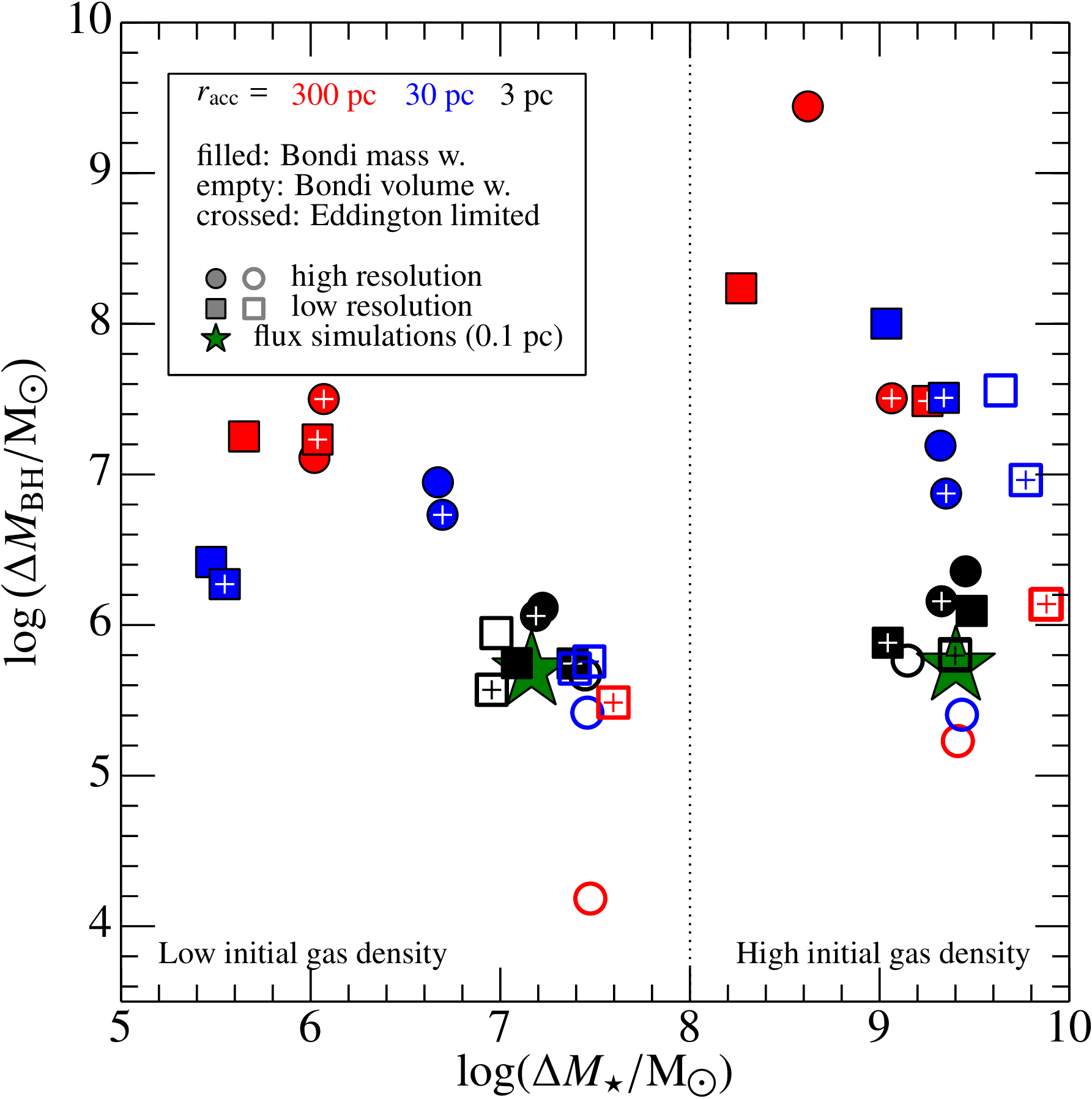}
\caption{Accreted BH mass vs stellar mass for all the simulations of Tables~\ref{tab1}-\ref{tabflux} and Table~\ref{tab3}. Symbols are the same as in Figure~\ref{fig:scatter2}.}
 \label{fig:scatter}
 \end{figure}

Finally, as expected, there is convergence between  Bondi and  flux accretion when the former is calculated using information in the immediate surroundings of the BH, at a distance of the scale of the Bondi radius. As shown in Figures~\ref{fig:summary} and \ref{fig:scatter}, the spread in BH and galaxy mass is within a factor of two for all runs with an accretion radius of 3~pc, regardless of the numerical implementation and of the galaxy gas content.

%
%
%
\subsection{Comparison with previous works}
Our results can be compared with other recent papers on BH accretion and feedback using high resolution simulations. Spherical Bondi accretion on supermassive BH has been studied in \citet{barai+11}, by performing high resolution simulations with the 3D code \textsc{gadget-3}. The spatial scales involved range from 0.1 to 20-200~pc, with a softening of 0.005-0.02~pc (gas self-gravity is not considered), thus the Bondi radius is resolved. Compton heating and cooling are included, but momentum feedback is not. The BH accretes all particles crossing the minimum radius of the simulation and the BH luminosity is artificially kept constant during the evolution.
While \citet{barai+11} clearly show that an SPH code can follow the steady evolution of an adiabatic Bondi solution without major issues, in presence of radiative heating and cooling the picture is modified. At intermediate $\Lbh$ non spherical cold clumps can be formed via thermal instability, while at high $\Lbh$ an outflow is created, reducing the accretion rate of a factor of $10^3$ (see their Figure~13).

The formation of a multiphase gas is further investigated in \citet{barai+12} using the same setup of the previous work. They confirm that formation of cold gas clumps in a spherical setup is a consequence of thermal instabilities acting on small perturbations of the initial conditions. These cold filaments accrete on the BH and trigger large fluctuations on short timescales in the accretion rate (like our reference model). The mass inflow rate ratio between the cold and the hot phase is calculated to constrain the boost parameter $\alpha$ commonly employed in large scale simulations. They find a value $\leq 1-5$ at the Bondi radius, much lower respect the usual factor $\alpha=100$ used in cosmological simulations, meaning that they do not find an important unresolved fraction of gas hidden in the cold phase. Finally, they argue that $\alpha$ can be resolution dependent. We confirm their results and extend them in the case of more realistic AGN feedback, galaxy model and in presence of star formation. The additional feedback mechanisms in our work are able to produce a multiphase gas in almost every circumstance, while they obtain a multiphase medium and a non-smooth BH accretion rate only in case of high BH luminosity. In addition, we find that the Bondi rate is not a good proxy for accretion when the Bondi radius is not resolved.


A comparable work regarding explored spatial scales and resolutions but with a more realistic feedback  is presented in \citet{hopkins.etal2016}. They perform 3D simulations of the first 100 pc around an isolated BH with the GIZMO code, assuming a force softening of $\epsilon=0.02$~pc, including gas self-gravity, rotation, radiative cooling down to 10~K and star formation. The compton heating/cooling and AGN feedback are the same of the present work\footnote{The adopted AGN feedback efficiencies are higher in \citet{hopkins.etal2016}. For our models $\eta=\Mout/\Mdotbh=0.4$ and $\vw=10^{4}~\kms$, while \citet{hopkins.etal2016} adopt much higher values of $\eta$ (up to 60, named $\beta$ in their Table~1), and in general lower outflow velocities ($500-5000~\kms$).}. Moreover, the BH accretes immediately any gas particle having an apocentric radius $<2.8\epsilon$. They compare this accretion scheme (that can represent the meshless counterpart of the flux accretion in grid codes) with two different types of Bondi approaches, namely $\dot{M}=4\upi G^2 \Mbh^2 \rho_\mathrm{gas}\cs^{-3}$ and $\dot{M}=4\upi G^2 M_\mathrm{enc}^2(<R) \rho_\mathrm{gas}(\cs^2+\langle V_\mathrm{gas-bh}^2 \rangle)^{-3/2}$, where the latter takes into account the mass outside the BH as a source of gravity for the gas outside the centre of the galaxy (however, it is not made explicit  how $\rho_\mathrm{gas}$ and $\cs$ are calculated). Both schemes fail to reproduce the correct inflow rate in simulations without AGN feedback by overestimating the BH accretion rate of a factor of $10^8$ and $10^3$, respectively \citep[see Figure~6 in][note that the Bondi accretion is computed, but not used to calculate feedback]{hopkins.etal2016}. The accretion of cold and molecular gas is regulated by gravitational torques in gaseous disc, while Bondi accretion assumes a spherical gas distribution with no rotation, thus the low $\cs$ of the rotating gas produces a large accretion rate in the Bondi formalism.

A direct comparison between the different models in \citet{hopkins.etal2016} with our results is not straightforward, due to the low \AN{rotation regime in our 2.5D simulations (see Sec.~\ref{initial_conditions})}. However, from the point of view of BH accretion, these two studies are  complementary: accretion in \citet{hopkins.etal2016} is completely regulated by the gravitational toques of the gas+stars disc, thus confirming and quantifying the inadequacy of the Bondi accretion in presence of rotation. On the other hand, we show that the Bondi accretion does not predict the right accretion rate even in absence of angular momentum. This can depend on resolution, feedback, volume and algorithm employed to calculate density and sound speed in the Bondi formula, producing both an over- and under-estimate of the accretion rate. When BH winds are activated, \citet{hopkins.etal2016} find that AGN feedback can evacuate the polar regions with high velocity outflows (5 -- $30\e{3}~\kms$), suppressing star formation and producing a BH accretion rate with an extreme time variability, in the same way mechanical feedback in our reference model is able to clear the innermost 100~pc from cold gas, thus producing an extremely fluctuating BH accretion rate that is maintained below the Eddington limit in a self-consistent way.

\citet{curtis.etal2015} attempt to bridge the scale of the Bondi radius to the resolutions typical of galactic scale simulations with a new super-Lagrangian refinement in the Voronoi-mesh AREPO code. The cells inside the smoothing length of the BH are refined almost down to the Bondi radius, and the Bondi rate is then computed from the mass weighted mean of density and sound speed over the cells. They apply this novel refinement scheme to 3D simulations of an isolated Milky-way like object, by employing different types of feedback (thermal and/or momentum, both isotropic and non-isotropic). The simulations with super-Lagrangian refinement show somewhat lower rates of gas accretion than those without (note that this resolution test is performed with isotropic thermal feedback only), due to the higher temperature generated by feedback in the higher resolution run. While this result agrees with our work, the difference in the final BH mass between different refinements found in \citet{curtis.etal2015} is small. We designed two tailored simulations with a setup similar to \cite{curtis.etal2015}. We adopted their accretion model with $\racc = 700~\pc$, a 1~pc resolution grid (see their Fig.~7) and $n_0 = 0.01~\cc$, $\Mbhi=10^6~\Msun$. The two runs employ a mass weighted Bondi accretion and a flux accretion, respectively. The results are in line with our main set of high resolution simulations presented in Section~\ref{sec:bondi_high_res}. The mass weighted Bondi accretion run overestimates the BH accreted mass with respect to the flux accretion at the same resolution and at 0.1~pc resolution by a factor $\simeq 5$ and 25, respectively. The BH in the Bondi run accretes at Eddington rate until  feedback is able to evacuate the accretion radius from cold gas, therefore its evolution is quite smooth, in contrast to the simulations using flux accretion. We find that the main trends resemble the Bondi mass weighted high resolution runs presented in Section~\ref{sec:bondi_high_res}, where the BH accretion is always overestimated when compared to the flux accretion.  This is due to the poor feedback efficiency in stopping the accretion, while in the flux accretion runs the BH feeding is much more irregular, and feedback is able to halt  accretion more effectively. This can explain in part the BH mass rapid increase in Figure~12 of \citet{curtis.etal2015} when momentum feedback only is employed. The scenario where AGN feedback is not effective in halting the accretion at large $\racc$, due to the large amount of cold gas inside the accretion radius, is thus confirmed. The super-Lagrangian refinement presented in \citet{curtis.etal2015} is promising, given its ability to dynamically resolve the region around BHs, and indeed in such a scheme the increased resolution can be used directly to apply a flux accretion to overcome the  issues of Bondi accretion.

\section{Conclusions}\label{sec:conclusions}
We employ 2.5D hydrodynamical simulations of an isolated galaxy to test when and whether Bondi accretion is a good proxy for BH feeding. In cosmological simulations the cold gas phase and the Bondi radius are not resolved, thus the ISM temperature is expected to be overestimated and its density underestimated. To correct for these effects,  a boost parameter $\alpha$ is often employed to increase the BH accretion rate. The large amount of sub-grid recipes based on Bondi accretion present in the literature motivated us to study how well BH accretion and the subsequent AGN feedback on galactic scales can be captured.

We include AGN radiative and mechanical feedback together with a 0.1~pc resolution grid. Our simulation setup consists of an isolated galaxy with a numerical grid (in the reference case) spanning from 0.1~pc to 250~kpc, thus resolving the Bondi radius for all the temperatures. Our reference runs implement flux accretion on the innermost grid. We compare the reference case with simulations employing Bondi accretion at different resolutions, and with different tweaks of the implementation (mass vs volume weighting). The main results are summarized below:

(1) The reference runs show a self-regulated accretion that never exceeds the Eddington limit, without any capping explicitly included. The hydrodynamics is characterized by irregular and highly variable cycles of cooling, accretion, AGN feedback.

(2) When mass weighting  is employed to estimate gas density and sound speed in the Bondi formalism in high resolution simulations (where the grid still resolves the Bondi radius), the BH accreted mass is overestimated up to a factor of 100. This is due to the complex interplay between feeding and feedback. The feedback efficiency in clearing the BH surroundings is a function of the accretion radius. At large accretion radii the AGN feedback is less effective in evacuating the accretion region, since the gas mass accumulated is too large to be swept away or heated up, leading to a smoother accretion and in the end to a larger BH mass. 

(3) In simulations with progressively lower resolution \AN{and larger $\racc$}, in absence of AGN feedback, the ISM temperature is overestimated and its density underestimated. However, the introduction of feedback \AN{in simulations with different resolutions} leads to an overestimate of the BH accretion due to poor feedback efficiency at \AN{lower resolutions}. Capping the accretion rate at the Eddington limit only enhances this trend since it does not allow the BH to reach the luminosity needed to effectively halt accretion.

(4) \AN{Volume weighted Bondi accretion underestimates the BH accreted mass when the high resolution grid is employed.} BH accretion is dominated by hot gas, as a result of the volume weighted averaging. In this scenario, the use of a boost parameter is justified, although the exact value of the boost factor is resolution and problem dependant.

(5) Unexpectedly, reducing the resolution in volume weighted Bondi simulations, BH and stellar mass are similar to the ``flux" case. This occurs because both the accretion rate and AGN feedback are underestimated, and compensate each other. 

(6) In general, we show that convergence between Bondi and flux accretion is reached only when the accretion radius resolves the Bondi radius.

\section*{Acknowledgements}
We thank Luca Ciotti and Jeremiah Ostriker for useful discussion and comments, Greg Novak for providing an initial version of
the code used here and the anonymous referee for helpful comments and contributions. AN thanks Tilman Hartwig, Muhammad Latif for interesting discussions, Rebekka Bieri and Mélanie Habouzit for help with the Horizon Cluster setup. AN and MV acknowledge funding from the European Research Council under the European Community’s Seventh Framework Programme (FP7/2007-2013 Grant Agreement no. 614199, project ``BLACK"). This work has made use of the Horizon Cluster funded by Institut d'Astrophysique de Paris. We thank Stephane Rouberol for running smoothly this cluster for us.

\bibliographystyle{mn2e}
\bibliography{citations.bib}

\begin{thebibliography}{97}
\expandafter\ifx\csname natexlab\endcsname\relax\def\natexlab#1{#1}\fi

\bibitem[{{Angl{\'e}s-Alc{\'a}zar}
  {et~al}\mbox{.}(2016){Angl{\'e}s-Alc{\'a}zar}, {Dav{\'e}},
  {Faucher-Gigu{\`e}re}, {{\"O}zel}, \& {Hopkins}}]{angles-Alcazar.etal2016}
{Angl{\'e}s-Alc{\'a}zar} D., {Dav{\'e}} R., {Faucher-Gigu{\`e}re} C.-A.,
  {{\"O}zel} F., {Hopkins} P.~F., 2016, preprint (arXiv:1603.08007)

\bibitem[{{Barai} {et~al}\mbox{.}(2011){Barai}, {Proga}, \&
  {Nagamine}}]{barai+11}
{Barai} P., {Proga} D., {Nagamine} K., 2011, \mnras, 418, 591

\bibitem[{{Barai} {et~al}\mbox{.}(2012){Barai}, {Proga}, \&
  {Nagamine}}]{barai+12}
{Barai} P., {Proga} D., {Nagamine} K., 2012, \mnras, 424, 728

\bibitem[{{Barai} {et~al}\mbox{.}(2014){Barai}, {Viel}, {Murante}, {Gaspari},
  \& {Borgani}}]{barai.etal2014}
{Barai} P., {Viel} M., {Murante} G., {Gaspari} M., {Borgani} S., 2014, \mnras,
  437, 1456

\bibitem[{{Bellovary} {et~al}\mbox{.}(2010){Bellovary}, {Governato}, {Quinn},
  {Wadsley}, {Shen}, \& {Volonteri}}]{bellovary.etal2010}
{Bellovary} J.~M., {Governato} F., {Quinn} T.~R., {Wadsley} J., {Shen} S.,
  {Volonteri} M., 2010, ApJL, 721, L148

\bibitem[{{Bertin} \& {Lodato}(2001)}]{bertin.lodato.2001}
{Bertin} G., {Lodato} G., 2001, \aap, 370, 342

\bibitem[{{Bleuler} \& {Teyssier}(2014)}]{bleuler.teyssier2014}
{Bleuler} A., {Teyssier} R., 2014, \mnras, 445, 4015

\bibitem[{{Bondi}(1952)}]{Bondi1952}
{Bondi} H., 1952, \mnras, 112, 195

\bibitem[{{Bondi} \& {Hoyle}(1944)}]{bondi.hoyle1944}
{Bondi} H., {Hoyle} F., 1944, \mnras, 104, 273

\bibitem[{{Booth} \& {Schaye}(2009)}]{booth.schaye2009}
{Booth} C.~M., {Schaye} J., 2009, \mnras, 398, 53

\bibitem[{{Bryan} {et~al}\mbox{.}(2014){Bryan}, {Norman}, {O'Shea}, {Abel},
  {Wise}, {Turk}, {Reynolds}, {Collins}, {Wang}, {Skillman}, {Smith},
  {Harkness}, {Bordner}, {Kim}, {Kuhlen}, {Xu}, {Goldbaum}, {Hummels},
  {Kritsuk}, {Tasker}, {Skory}, {Simpson}, {Hahn}, {Oishi}, {So}, {Zhao},
  {Cen}, {Li}, \& {Enzo Collaboration}}]{bryan.etal2014}
{Bryan} G.~L. {et~al.}, 2014, \apjs, 211, 19

\bibitem[{{Choi} {et~al}\mbox{.}(2012){Choi}, {Ostriker}, {Naab}, \&
  {Johansson}}]{choi.etal2012}
{Choi} E., {Ostriker} J.~P., {Naab} T., {Johansson} P.~H., 2012, \apj, 754, 125

\bibitem[{{Choi} {et~al}\mbox{.}(2015){Choi}, {Ostriker}, {Naab}, {Oser}, \&
  {Moster}}]{choi.etal2015}
{Choi} E., {Ostriker} J.~P., {Naab} T., {Oser} L., {Moster} B.~P., 2015,
  \mnras, 449, 4105

\bibitem[{{Ciotti} {et~al}\mbox{.}(2009){Ciotti}, {Morganti}, \& {de
  Zeeuw}}]{ciotti.etal2009b}
{Ciotti} L., {Morganti} L., {de Zeeuw} P.~T., 2009, \mnras, 393, 491

\bibitem[{{Ciotti} \& {Ostriker}(1997)}]{ciotti.ostriker1997}
{Ciotti} L., {Ostriker} J.~P., 1997, \apjl, 487, L105

\bibitem[{{Ciotti} \& {Ostriker}(2001)}]{ciotti.ostriker2001}
{Ciotti} L., {Ostriker} J.~P., 2001, \apj, 551, 131

\bibitem[{{Ciotti} \& {Ostriker}(2007)}]{ciotti.ostriker2007}
{Ciotti} L., {Ostriker} J.~P., 2007, \apj, 665, 1038

\bibitem[{{Ciotti} \& {Ostriker}(2012)}]{ciotti.Ostriker.2012}
{Ciotti} L., {Ostriker} J.~P., 2012, in Astrophysics and Space Science Library,
  Vol. 378, Astrophysics and Space Science Library, {Kim} D.-W., {Pellegrini}
  S., eds., p.~83

\bibitem[{{Ciotti} {et~al}\mbox{.}(2016){Ciotti}, {Pellegrini}, {Negri}, \&
  {Ostriker}}]{ciotti.etal2016}
{Ciotti} L., {Pellegrini} S., {Negri} A., {Ostriker} J.~P., 2016, preprint
  (arXiv:1608.03403)

\bibitem[{{Curtis} \& {Sijacki}(2015)}]{curtis.etal2015}
{Curtis} M., {Sijacki} D., 2015, \mnras, 454, 3445

\bibitem[{{Curtis} \& {Sijacki}(2016)}]{curtis.sijacki2016}
{Curtis} M., {Sijacki} D., 2016, \mnras, 463, 63

\bibitem[{{Debuhr} {et~al}\mbox{.}(2011){Debuhr}, {Quataert}, \&
  {Ma}}]{debhur.etal2011}
{Debuhr} J., {Quataert} E., {Ma} C.-P., 2011, \mnras, 412, 1341

\bibitem[{{Debuhr} {et~al}\mbox{.}(2010){Debuhr}, {Quataert}, {Ma}, \&
  {Hopkins}}]{debuhr.etal2010}
{Debuhr} J., {Quataert} E., {Ma} C.-P., {Hopkins} P., 2010, \mnras, 406, L55

\bibitem[{{DeGraf} {et~al}\mbox{.}(2012){DeGraf}, {Di Matteo}, {Khandai},
  {Croft}, {Lopez}, \& {Springel}}]{degraf.etal2012}
{DeGraf} C., {Di Matteo} T., {Khandai} N., {Croft} R., {Lopez} J., {Springel}
  V., 2012, \mnras, 424, 1892

\bibitem[{{Degraf} {et~al}\mbox{.}(2011){Degraf}, {Di Matteo}, \&
  {Springel}}]{degraf.etal2011}
{Degraf} C., {Di Matteo} T., {Springel} V., 2011, \mnras, 413, 1383

\bibitem[{{Di Matteo} {et~al}\mbox{.}(2008){Di Matteo}, {Colberg}, {Springel},
  {Hernquist}, \& {Sijacki}}]{dimatteo.etal2008}
{Di Matteo} T., {Colberg} J., {Springel} V., {Hernquist} L., {Sijacki} D.,
  2008, \apj, 676, 33

\bibitem[{{Di Matteo} {et~al}\mbox{.}(2016){Di Matteo}, {Croft}, {Feng},
  {Waters}, \& {Wilkins}}]{dimatteo.etal2016}
{Di Matteo} T., {Croft} R.~A.~C., {Feng} Y., {Waters} D., {Wilkins} S., 2016,
  preprint (arXiv:1606.08871)

\bibitem[{{Di Matteo} {et~al}\mbox{.}(2005){Di Matteo}, {Springel}, \&
  {Hernquist}}]{dimatteo.etal2005}
{Di Matteo} T., {Springel} V., {Hernquist} L., 2005, \nat, 433, 604

\bibitem[{{Dubois} {et~al}\mbox{.}(2012){Dubois}, {Devriendt}, {Slyz}, \&
  {Teyssier}}]{dubois.etal2012}
{Dubois} Y., {Devriendt} J., {Slyz} A., {Teyssier} R., 2012, \mnras, 420, 2662

\bibitem[{{Dubois} {et~al}\mbox{.}(2014){Dubois}, {Pichon}, {Welker}, {Le
  Borgne}, {Devriendt}, {Laigle}, {Codis}, {Pogosyan}, {Arnouts}, {Benabed},
  {Bertin}, {Blaizot}, {Bouchet}, {Cardoso}, {Colombi}, {de Lapparent},
  {Desjacques}, {Gavazzi}, {Kassin}, {Kimm}, {McCracken}, {Milliard},
  {Peirani}, {Prunet}, {Rouberol}, {Silk}, {Slyz}, {Sousbie}, {Teyssier},
  {Tresse}, {Treyer}, {Vibert}, \& {Volonteri}}]{dubois.etal2014}
{Dubois} Y. {et~al.}, 2014, \mnras, 444, 1453

\bibitem[{{Elahi} {et~al}\mbox{.}(2016){Elahi}, {Knebe}, {Pearce}, {Power},
  {Yepes}, {Cui}, {Cunnama}, {Kay}, {Sembolini}, {Beck}, {Dav{\'e}},
  {February}, {Huang}, {Katz}, {McCarthy}, {Murante}, {Perret}, {Puchwein},
  {Saro}, \& {Teyssier}}]{elahi.etal2016}
{Elahi} P.~J. {et~al.}, 2016, \mnras, 458, 1096

\bibitem[{{Fabian}(1999)}]{fabian1999}
{Fabian} A.~C., 1999, \mnras, 308, L39

\bibitem[{{Fabian}(2012)}]{fabian2012}
{Fabian} A.~C., 2012, \araa, 50, 455

\bibitem[{{Feng} {et~al}\mbox{.}(2016){Feng}, {Di-Matteo}, {Croft}, {Bird},
  {Battaglia}, \& {Wilkins}}]{feng.etal2016}
{Feng} Y., {Di-Matteo} T., {Croft} R.~A., {Bird} S., {Battaglia} N., {Wilkins}
  S., 2016, \mnras, 455, 2778

\bibitem[{{Ferrarese} \& {Merritt}(2000)}]{ferrarese.merrit2000}
{Ferrarese} L., {Merritt} D., 2000, \apjl, 539, L9

\bibitem[{{Gaspari} {et~al}\mbox{.}(2015){Gaspari}, {Brighenti}, \&
  {Temi}}]{gaspari.etal2015}
{Gaspari} M., {Brighenti} F., {Temi} P., 2015, \aap, 579, A62

\bibitem[{{Gaspari} {et~al}\mbox{.}(2013){Gaspari}, {Ruszkowski}, \&
  {Oh}}]{gaspari.etal2013}
{Gaspari} M., {Ruszkowski} M., {Oh} S.~P., 2013, \mnras, 432, 3401

\bibitem[{{Gaspari} {et~al}\mbox{.}(2016){Gaspari}, {Temi}, \&
  {Brighenti}}]{gaspari.etal2016}
{Gaspari} M., {Temi} P., {Brighenti} F., 2016, preprint (arXiv:1608.08216)

\bibitem[{{Gebhardt} {et~al}\mbox{.}(2000){Gebhardt}, {Bender}, {Bower},
  {Dressler}, {Faber}, {Filippenko}, {Green}, {Grillmair}, {Ho}, {Kormendy},
  {Lauer}, {Magorrian}, {Pinkney}, {Richstone}, \&
  {Tremaine}}]{gebhardt.etal2000}
{Gebhardt} K. {et~al.}, 2000, \apjl, 539, L13

\bibitem[{{G{\"u}ltekin} {et~al}\mbox{.}(2009){G{\"u}ltekin}, {Richstone},
  {Gebhardt}, {Lauer}, {Tremaine}, {Aller}, {Bender}, {Dressler}, {Faber},
  {Filippenko}, {Green}, {Ho}, {Kormendy}, {Magorrian}, {Pinkney}, \&
  {Siopis}}]{gultekin.etal2009}
{G{\"u}ltekin} K. {et~al.}, 2009, \apj, 698, 198

\bibitem[{{Hayes} {et~al}\mbox{.}(2006){Hayes}, {Norman}, {Fiedler}, {Bordner},
  {Li}, {Clark}, {ud-Doula}, \& {Mac Low}}]{hayesetal2006}
{Hayes} J.~C., {Norman} M.~L., {Fiedler} R.~A., {Bordner} J.~O., {Li} P.~S.,
  {Clark} S.~E., {ud-Doula} A., {Mac Low} M.-M., 2006, \apjs, 165, 188

\bibitem[{{Hensley} {et~al}\mbox{.}(2014){Hensley}, {Ostriker}, \&
  {Ciotti}}]{hensley.etal2014}
{Hensley} B.~S., {Ostriker} J.~P., {Ciotti} L., 2014, \apj, 789, 78

\bibitem[{{Hirschmann} {et~al}\mbox{.}(2014){Hirschmann}, {Dolag}, {Saro},
  {Bachmann}, {Borgani}, \& {Burkert}}]{hirschmann.etal2014}
{Hirschmann} M., {Dolag} K., {Saro} A., {Bachmann} L., {Borgani} S., {Burkert}
  A., 2014, \mnras, 442, 2304

\bibitem[{{Hopkins} \& {Quataert}(2010)}]{hopkins.quataert2010}
{Hopkins} P.~F., {Quataert} E., 2010, \mnras, 407, 1529

\bibitem[{{Hopkins} \& {Quataert}(2011)}]{hopkins.quataert2011}
{Hopkins} P.~F., {Quataert} E., 2011, \mnras, 415, 1027

\bibitem[{{Hopkins} {et~al}\mbox{.}(2016){Hopkins}, {Torrey},
  {Faucher-Gigu{\`e}re}, {Quataert}, \& {Murray}}]{hopkins.etal2016}
{Hopkins} P.~F., {Torrey} P., {Faucher-Gigu{\`e}re} C.-A., {Quataert} E.,
  {Murray} N., 2016, \mnras, 458, 816

\bibitem[{{Hoyle} \& {Lyttleton}(1939)}]{hoyle.lyttleton1939}
{Hoyle} F., {Lyttleton} R.~A., 1939, Proceedings of the Cambridge Philosophical
  Society, 35, 405

\bibitem[{{Inayoshi} {et~al}\mbox{.}(2016){Inayoshi}, {Haiman}, \&
  {Ostriker}}]{inayoshi.etal2015}
{Inayoshi} K., {Haiman} Z., {Ostriker} J.~P., 2016, \mnras, 459, 3738

\bibitem[{{Jaffe}(1983)}]{jaffe1983}
{Jaffe} W., 1983, \mnras, 202, 995

\bibitem[{{Kennicutt}(1998)}]{kennicutt1998}
{Kennicutt}, Jr. R.~C., 1998, \apj, 498, 541

\bibitem[{{Khandai} {et~al}\mbox{.}(2015){Khandai}, {Di Matteo}, {Croft},
  {Wilkins}, {Feng}, {Tucker}, {DeGraf}, \& {Liu}}]{khandai.etal2015}
{Khandai} N., {Di Matteo} T., {Croft} R., {Wilkins} S., {Feng} Y., {Tucker} E.,
  {DeGraf} C., {Liu} M.-S., 2015, \mnras, 450, 1349

\bibitem[{{Kim} {et~al}\mbox{.}(2011){Kim}, {Wise}, {Alvarez}, \&
  {Abel}}]{kim.etal2011}
{Kim} J.-h., {Wise} J.~H., {Alvarez} M.~A., {Abel} T., 2011, \apj, 738, 54

\bibitem[{{King}(2005)}]{king2005}
{King} A., 2005, \apjl, 635, L121

\bibitem[{{Kormendy}(2016)}]{kormendy2016}
{Kormendy} J., 2016, Galactic Bulges, 418, 431

\bibitem[{{Kormendy} \& {Ho}(2013)}]{kormendy.ho2013}
{Kormendy} J., {Ho} L.~C., 2013, \araa, 51, 511

\bibitem[{{Korol} {et~al}\mbox{.}(2016){Korol}, {Ciotti}, \&
  {Pellegrini}}]{korol.etal2016}
{Korol} V., {Ciotti} L., {Pellegrini} S., 2016, \mnras, 460, 1188

\bibitem[{{Kurosawa} \& {Proga}(2009{\natexlab{a}})}]{kurosawa.proga2009a}
{Kurosawa} R., {Proga} D., 2009{\natexlab{a}}, \mnras, 397, 1791

\bibitem[{{Kurosawa} \& {Proga}(2009{\natexlab{b}})}]{kurosawa.proga2009b}
{Kurosawa} R., {Proga} D., 2009{\natexlab{b}}, \apj, 693, 1929

\bibitem[{{Kurosawa} {et~al}\mbox{.}(2009){Kurosawa}, {Proga}, \&
  {Nagamine}}]{kurosawa.etal2009}
{Kurosawa} R., {Proga} D., {Nagamine} K., 2009, \apj, 707, 823

\bibitem[{{Le Brun} {et~al}\mbox{.}(2014){Le Brun}, {McCarthy}, {Schaye}, \&
  {Ponman}}]{lebrun.etal2014}
{Le Brun} A.~M.~C., {McCarthy} I.~G., {Schaye} J., {Ponman} T.~J., 2014,
  \mnras, 441, 1270

\bibitem[{{Li} {et~al}\mbox{.}(2013){Li}, {Ostriker}, \&
  {Sunyaev}}]{li.etal2013}
{Li} J., {Ostriker} J., {Sunyaev} R., 2013, \apj, 767, 105

\bibitem[{{Lusso} \& {Ciotti}(2011)}]{lusso.ciotti2011}
{Lusso} E., {Ciotti} L., 2011, \aap, 525, A115

\bibitem[{{Magorrian} {et~al}\mbox{.}(1998){Magorrian}, {Tremaine},
  {Richstone}, {Bender}, {Bower}, {Dressler}, {Faber}, {Gebhardt}, {Green},
  {Grillmair}, {Kormendy}, \& {Lauer}}]{magorrian1998}
{Magorrian} J. {et~al.}, 1998, \aj, 115, 2285

\bibitem[{{McCarthy} {et~al}\mbox{.}(2016){McCarthy}, {Schaye}, {Bird}, \& {Le
  Brun}}]{mcCarthy.etal2016}
{McCarthy} I.~G., {Schaye} J., {Bird} S., {Le Brun} A.~M.~C., 2016, preprint
  (arXiv:1603.02702)

\bibitem[{{McConnell} \& {Ma}(2013)}]{mcconnell.ma2013}
{McConnell} N.~J., {Ma} C.-P., 2013, \apj, 764, 184

\bibitem[{{Narayan} \& {Yi}(1994)}]{narayan.yi1994}
{Narayan} R., {Yi} I., 1994, \apjl, 428, L13

\bibitem[{{Negri} {et~al}\mbox{.}(2014{\natexlab{a}}){Negri}, {Ciotti}, \&
  {Pellegrini}}]{negri.etal2014}
{Negri} A., {Ciotti} L., {Pellegrini} S., 2014{\natexlab{a}}, \mnras, 439, 823

\bibitem[{{Negri} {et~al}\mbox{.}(2015){Negri}, {Pellegrini}, \&
  {Ciotti}}]{negri.etal2015}
{Negri} A., {Pellegrini} S., {Ciotti} L., 2015, \mnras, 451, 1212

\bibitem[{{Negri} {et~al}\mbox{.}(2014{\natexlab{b}}){Negri}, {Posacki},
  {Pellegrini}, \& {Ciotti}}]{negri.etal2014b}
{Negri} A., {Posacki} S., {Pellegrini} S., {Ciotti} L., 2014{\natexlab{b}},
  \mnras, 445, 1351

\bibitem[{{Novak} {et~al}\mbox{.}(2011){Novak}, {Ostriker}, \&
  {Ciotti}}]{novak.etal.2011}
{Novak} G.~S., {Ostriker} J.~P., {Ciotti} L., 2011, \apj, 737, 26

\bibitem[{{Novak} {et~al}\mbox{.}(2012){Novak}, {Ostriker}, \&
  {Ciotti}}]{novak.etal2012}
{Novak} G.~S., {Ostriker} J.~P., {Ciotti} L., 2012, \mnras, 427, 2734

\bibitem[{{Ostriker} {et~al}\mbox{.}(2010){Ostriker}, {Choi}, {Ciotti},
  {Novak}, \& {Proga}}]{ostriker.etal2010}
{Ostriker} J.~P., {Choi} E., {Ciotti} L., {Novak} G.~S., {Proga} D., 2010,
  \apj, 722, 642

\bibitem[{{Park} \& {Ricotti}(2012)}]{park.ricotti2012}
{Park} K., {Ricotti} M., 2012, \apj, 747, 9

\bibitem[{{Pelupessy} {et~al}\mbox{.}(2007){Pelupessy}, {Di Matteo}, \&
  {Ciardi}}]{pelupessy.etal2007}
{Pelupessy} F.~I., {Di Matteo} T., {Ciardi} B., 2007, \apj, 665, 107

\bibitem[{{Power} {et~al}\mbox{.}(2011){Power}, {Nayakshin}, \&
  {King}}]{power.etal2011}
{Power} C., {Nayakshin} S., {King} A., 2011, \mnras, 412, 269

\bibitem[{{Prasad} {et~al}\mbox{.}(2015){Prasad}, {Sharma}, \&
  {Babul}}]{prasad.etal.2015}
{Prasad} D., {Sharma} P., {Babul} A., 2015, \apj, 811, 108

\bibitem[{{Robertson} {et~al}\mbox{.}(2006){Robertson}, {Bullock}, {Cox}, {Di
  Matteo}, {Hernquist}, {Springel}, \& {Yoshida}}]{robertson.etal2006}
{Robertson} B., {Bullock} J.~S., {Cox} T.~J., {Di Matteo} T., {Hernquist} L.,
  {Springel} V., {Yoshida} N., 2006, \apj, 645, 986

\bibitem[{{Rosas-Guevara} {et~al}\mbox{.}(2015){Rosas-Guevara}, {Bower},
  {Schaye}, {Furlong}, {Frenk}, {Booth}, {Crain}, {Dalla Vecchia}, {Schaller},
  \& {Theuns}}]{rosas-Guevara.etal2015}
{Rosas-Guevara} Y.~M. {et~al.}, 2015, \mnras, 454, 1038

\bibitem[{{Schaye} {et~al}\mbox{.}(2015){Schaye}, {Crain}, {Bower}, {Furlong},
  {Schaller}, {Theuns}, {Dalla Vecchia}, {Frenk}, {McCarthy}, {Helly},
  {Jenkins}, {Rosas-Guevara}, {White}, {Baes}, {Booth}, {Camps}, {Navarro},
  {Qu}, {Rahmati}, {Sawala}, {Thomas}, \& {Trayford}}]{schaye.etal2015}
{Schaye} J. {et~al.}, 2015, \mnras, 446, 521

\bibitem[{{Schaye} {et~al}\mbox{.}(2010){Schaye}, {Dalla Vecchia}, {Booth},
  {Wiersma}, {Theuns}, {Haas}, {Bertone}, {Duffy}, {McCarthy}, \& {van de
  Voort}}]{schaye.etal2010}
{Schaye} J. {et~al.}, 2010, \mnras, 402, 1536

\bibitem[{{Shakura} \& {Sunyaev}(1973)}]{Shakura.Sunyaev73}
{Shakura} N.~I., {Sunyaev} R.~A., 1973, \aap, 24, 337

\bibitem[{{Sijacki} {et~al}\mbox{.}(2007){Sijacki}, {Springel}, {Di Matteo}, \&
  {Hernquist}}]{sijacki.etal2007}
{Sijacki} D., {Springel} V., {Di Matteo} T., {Hernquist} L., 2007, \mnras, 380,
  877

\bibitem[{{Sijacki} {et~al}\mbox{.}(2015){Sijacki}, {Vogelsberger}, {Genel},
  {Springel}, {Torrey}, {Snyder}, {Nelson}, \& {Hernquist}}]{sijacki.etal2015}
{Sijacki} D., {Vogelsberger} M., {Genel} S., {Springel} V., {Torrey} P.,
  {Snyder} G.~F., {Nelson} D., {Hernquist} L., 2015, \mnras, 452, 575

\bibitem[{{Silk} \& {Rees}(1998)}]{silk.rees1998}
{Silk} J., {Rees} M.~J., 1998, \aap, 331, L1

\bibitem[{{Springel}(2010)}]{springel2010}
{Springel} V., 2010, \mnras, 401, 791

\bibitem[{{Springel} {et~al}\mbox{.}(2005){Springel}, {Di Matteo}, \&
  {Hernquist}}]{springel.etal2005}
{Springel} V., {Di Matteo} T., {Hernquist} L., 2005, \mnras, 361, 776

\bibitem[{{Steinborn} {et~al}\mbox{.}(2015){Steinborn}, {Dolag}, {Hirschmann},
  {Prieto}, \& {Remus}}]{steinborn.etal2015}
{Steinborn} L.~K., {Dolag} K., {Hirschmann} M., {Prieto} M.~A., {Remus} R.-S.,
  2015, \mnras, 448, 1504

\bibitem[{{Stone} \& {Norman}(1992)}]{stone.norman.1992}
{Stone} J.~M., {Norman} M.~L., 1992, \apjs, 80, 753

\bibitem[{{Teyssier}(2002)}]{teyssier2002}
{Teyssier} R., 2002, \aap, 385, 337

\bibitem[{{Teyssier} {et~al}\mbox{.}(2011){Teyssier}, {Moore}, {Martizzi},
  {Dubois}, \& {Mayer}}]{teyssier.etal2011}
{Teyssier} R., {Moore} B., {Martizzi} D., {Dubois} Y., {Mayer} L., 2011,
  \mnras, 414, 195

\bibitem[{{Tremmel} {et~al}\mbox{.}(2016){Tremmel}, {Karcher}, {Governato},
  {Volonteri}, {Quinn}, {Pontzen}, \& {Anderson}}]{tremmel.etal2016}
{Tremmel} M., {Karcher} M., {Governato} F., {Volonteri} M., {Quinn} T.,
  {Pontzen} A., {Anderson} L., 2016, preprint (arXiv:1607.02151)

\bibitem[{{Vogelsberger} {et~al}\mbox{.}(2013){Vogelsberger}, {Genel},
  {Sijacki}, {Torrey}, {Springel}, \& {Hernquist}}]{vogelsberger.etal2013}
{Vogelsberger} M., {Genel} S., {Sijacki} D., {Torrey} P., {Springel} V.,
  {Hernquist} L., 2013, \mnras, 436, 3031

\bibitem[{{Vogelsberger} {et~al}\mbox{.}(2014{\natexlab{a}}){Vogelsberger},
  {Genel}, {Springel}, {Torrey}, {Sijacki}, {Xu}, {Snyder}, {Bird}, {Nelson},
  \& {Hernquist}}]{vogelsberger.etal2014b}
{Vogelsberger} M. {et~al.}, 2014{\natexlab{a}}, \nat, 509, 177

\bibitem[{{Vogelsberger} {et~al}\mbox{.}(2014{\natexlab{b}}){Vogelsberger},
  {Genel}, {Springel}, {Torrey}, {Sijacki}, {Xu}, {Snyder}, {Nelson}, \&
  {Hernquist}}]{vogelsberger.etal2014}
{Vogelsberger} M. {et~al.}, 2014{\natexlab{b}}, \mnras, 444, 1518

\bibitem[{{Volonteri} {et~al}\mbox{.}(2016){Volonteri}, {Dubois}, {Pichon}, \&
  {Devriendt}}]{volonteri.etal2016}
{Volonteri} M., {Dubois} Y., {Pichon} C., {Devriendt} J., 2016, \mnras, 460,
  2979

\bibitem[{{Volonteri} {et~al}\mbox{.}(2015){Volonteri}, {Silk}, \&
  {Dubus}}]{volonteri.etal2015}
{Volonteri} M., {Silk} J., {Dubus} G., 2015, \apj, 804, 148

\bibitem[{{Wurster} \& {Thacker}(2013)}]{wurster.thacker2013}
{Wurster} J., {Thacker} R.~J., 2013, \mnras, 431, 2513

\end{thebibliography}

\end{document}